\documentclass[twocolumn,aps,prd,showpacs,%
nofootinbib,eqsecnum]%
{revtex4}
\usepackage{graphicx}

\newcommand{\s}[1]{{\not{\! #1}}}

\begin{document}

\title{On the $S$-matrix renormalization in effective theories}

\author{K.~Semenov-Tian-Shansky$^{a,c}$}
\email[Email address: ]{K.Semenov@ulg.ac.be}
\author{A.~Vereshagin$^{b,c}$}
\email[Email address: ]{Alexander.Vereshagin@ift.uib.no}
\author{V.~Vereshagin$^c$}
\email[Email address: ]{vvv@av2467.spb.edu}
\affiliation{
$^a$Institut de Physique, B5a,
    Universit\'e de Li\`ege au Sart Tilman,
    B4000 Li\`ege 1, Belgium \\
$^b$Institute of Physics and Technology,
    Allegt.55,
    N5007 Bergen,
    Norway \\
$^c$Theor.\ Phys.\ Dept., Institute of Physics,
    St.Petersburg State University,
    St.Petersburg, Petrodvoretz,
    198504, Russia
}

\date{30 October 2005}

\begin{abstract}
This is the 5-th paper in the series devoted to explicit formulating
of the rules needed to manage an effective field theory of strong
interactions in
$S$-matrix sector. We discuss the principles of constructing the
meaningful perturbation series and formulate two basic ones:
uniformity and summability. Relying on these principles one obtains
the bootstrap conditions which restrict the allowed values of the
physical (observable) parameters appearing in the extended
perturbation scheme built for a given localizable effective theory.
The renormalization prescriptions needed to fix the finite parts of
counterterms in such a scheme can be divided into two subsets: minimal
--- needed to fix the S-matrix, and non-minimal --- for  eventual
calculation of Green functions; in this paper we consider only  the
minimal one. In particular, it is shown that in theories with the
amplitudes which asymptotic behavior is governed by known Regge
intercepts, the system of independent renormalization conditions only
contains those fixing the counterterm vertices with
$n \leq 3$ lines, while other prescriptions are determined by
self-consistency requirements. Moreover, the prescriptions for
$n \leq 3$ cannot be taken arbitrary: an infinite number of bootstrap
conditions should be respected. The concept of localizability,
introduced and explained in this article, is closely connected with
the notion of resonance in the framework of perturbative QFT. We
discuss this point and, finally, compare the corner stones of our
approach with the philosophy known as ``analytic
$S$-matrix''.
\end{abstract}

\pacs{02.30.Lt, 11.10.Gh, 11.10.Lm, 11.15.Bt, 11.80.-m}

\maketitle

\section{Introduction}
\label{sec-intro}
\mbox{}

This paper, together with the previous one
\cite{AVVV2}, is aimed to form the philosophical and theoretical base
for calculations proposed in
\cite{VVV}--\cite{POMI} and to outline the ways for further analysis,
therefore it is natural to review first the material presented in
those publications.

We are interested in constructing a self-consistent perturbation
technique for the infinite component effective field theories%
\footnote{
We use this term in its original meaning (see
\cite{WeinbEFT,WeinMONO})
with small modifications suggested in
\cite{AVVV2,POMI}.
Our definition is formulated in the beginning of
Sec.~\ref{sec-prelim},
the details are discussed in
Sec.~\ref{sec-localizability}.
}
of strong interactions. We work with Dyson's scheme, because it is the
only known way to combine Lorentz invariance, unitarity and cluster
decomposition principle with postulates of quantum mechanics. Thus,
the problems we have to do with include an infinite number of graphs
to be summed up at each loop order and the problem of ordering the
required renormalization conditions, since in such a theory one needs
to fix an infinite number of parameters to be able to calculate
amplitudes. In
\cite{AVVV1}
it has been shown that already the requirement of summability of tree
graphs --- the existence of well-defined tree level amplitudes ---
leads to strong limitations on the possible values of coupling
constants of a theory. However, in those articles some theoretical
statements, like
{\em meromorphy}
and
{\em polynomial boundedness}
of the tree level amplitude, were taken as postulates and only some
general arguments in their favor were given. With these assumptions it
became possible to obtain the system of
{\em bootstrap equations}
for masses and coupling constants of
$\pi\pi$
and
$\pi K$
resonances in nice agreement with experimental data. The main tool
used to derive them was the
{\em Cauchy expansion},
based on the celebrated Cauchy integral formula, which represents the
tree level amplitude as a well defined series in a given domain of the
space of kinematical variables.

In subsequent publications we fill some gaps left in the previous
analysis and discuss new concepts. Thus, in
\cite{POMI} (see also
\cite{MENU}) we suggest the notion of
{\em minimal parametrization} and in
\cite{AVVV2} the corresponding reduction theorem is proven. This
theorem explains why it is sufficient to consider only the minimal
(``on-shell-surviving'') vertices at each loop order of the
perturbation theory --- the fact implicitly used in
\cite{AVVV1} to parameterize amplitudes. Besides, in
\cite{POMI} we briefly discuss what we call the
{\em localizability} requirement --- the philosophy which, in
particular, serves as background for the requirements of polynomial
boundedness and meromorphy of tree level amplitudes. Since the last
point was not explained clear enough, we address it in this paper.

We start with summing up the main results of previous publication
\cite{AVVV2} in
Sec.~\ref{sec-prelim}. Continuing the logic line of that article we
consider the
{\em extended perturbation scheme} based on the interaction
Hamiltonian%
\footnote{
Throughout the paper when saying Hamiltonian we mean the Hamiltonian
density. Besides, we always imply that this density (in the
interaction picture) is written in the Lorentz-covariant form thus
using Wick's
$T$-product in Dyson's series; see
Sec.~\ref{sec-localizability} below.
}
which, along with the fields that describe true asymptotic states,
contains also auxiliary fields of fictitious unstable particles ---
resonances. In
Sec.~\ref{sec-basic} we formulate two mathematical principles:
{\em asymptotic uniformity} and
{\em summability}, which create a base for constructing the meaningful
perturbation series in such effective theory. Further, in
Sec.~\ref{sec-Cauchy} we briefly discuss the mathematical tool which
allows us to present the finite loop order amplitudes as convergent
functional series. This technique is exploited in
Secs.~\ref{sec-RP1}--\ref{sec-RP2}, where we demonstrate how the
requirement of crossing symmetry gives rise to the system of bootstrap
conditions and analyze the renormalization procedure. The results of
these two Sections are generalized in
Sec.~\ref{sec-RP3}, where phenomenological constraints are imposed
and the renormalization prescriptions are explicitly written; besides,
it is shown there that the bootstrap conditions obtained at any loop
order can be treated as the relations between
{\em physical observables}, which justify the legitimacy of our
preceding analysis of experimental data
\cite{AVVV1,MENU,Talks}.

The connection between the extended perturbation scheme and the
{\em localizable initial theory} (based on the Hamiltonian constructed
solely from the fields of true asymptotic states) is outlined in
Sec.~\ref{sec-localizability} which, along with
Sec.~\ref{sec-basic}, is the central one in this article. In
particular, we discuss here a (rather hypothetic) step-by-step
process of localization and explain our usage of the term
``{\em strong interaction}''. The localization process requires
introducing auxiliary free fields. The physical interpretation of
those fields is given in
Sec.~\ref{sec-resonances} where we discuss the meaning of the terms
``mass'' and ``width'' as the resonance classification parameters.

At last, the Section
\ref{sec-analyt} is devoted to comparative analysis of the effective
scattering theory philosophy with respect to that of analytic
$S$-matrix.

\section{Classification of the parameters: a brief review}
\label{sec-prelim}
\mbox{}

For the following discussion it is essential to recall the results
obtained in
\cite{AVVV2}. Referring to the analysis presented there we attempt to
be not too rigorous trying, instead, to make a picture clear.

We say that
{\em the field theory is effective if the interaction Hamiltonian in
the interaction picture contains all the monomials consistent with a
given algebraic (linear) symmetry}%
\footnote{
The original definition given in
\cite{WeinbEFT} employs Lagrangian. The reasons for this difference were
explained in
\cite{AVVV2} and
\cite{POMI}, see also
Sec.~\ref{sec-localizability} below. In general we do not imply any
other symmetry but Lorentz invariance, inclusion of any linear
internal symmetry being trivial. The dynamical (non-linear) symmetries
are briefly discussed in
\cite{AVVV2}.
}.
{\em Effective scattering} theory is just the effective field theory
only used to compute the
$S$-matrix, while the calculation of Green functions is not implied.
In particular, only the
$S$-matrix elements should be renormalized.

The fields of stable particles and resonances of arbitrary high spin
are present in effective Hamiltonian and every vertex is dotted by the
corresponding coupling constant. Therefore we are forced to deal with
an infinite set of parameters. By construction, the effective theory
is renormalizable but, to make use of this property, one needs an
infinite number of renormalization prescriptions. This looks
impractical until certain regularity reducing the number of
independent prescriptions, possibly up to some basic set, is found. As
shown in
\cite{AVVV2}, the
$S$-matrix in effective theory only depends upon a certain set of
parameters, which we called the
{\em resultant parameters} of various levels (or loop orders)
$l$. The value
$l=0$ labels the tree-level parameters,
$l=1$ --- one-loop level ones, and so on.

The parametrization implies the use of
{\em renormalized} perturbation theory (see, e.g.
\cite{Vasiliev2, Renorm, Peskin}),
so the interaction Hamiltonian is written as a sum of basic one plus
counterterms:
\[
{\cal H}_{\rm int} = {\cal H}_{\rm b} + {\cal H}_{\rm ct}\ .
\]
The coupling constants in
${\cal H}_{\rm b}$
are the physical ones, while each counterterm contributes starting
from appropriate loop order. The resultant parameters of order
$l=0$ include mass parameters%
\footnote{
Real masses, appearing in Feynman propagators. For stable particles
these are the physical (observable) masses --- see
Secs.~\ref{sec-RP3}--\ref{sec-resonances}.
}
and certain infinite polylinear combinations of the Hamiltonian
coupling constants with coefficients depending on masses. The
$l$-th level resultant parameters
${v}^{(l)}_{\cdots}$ with
$l \geq 1$ contain also the items depending on counterterm couplings
of orders
$l' \leq l$. In fact, it is the new counterterms arising at each new
loop order that makes this classification convenient in perturbative
calculations.

The construction of resultant parameters implies transition to the
{\em minimal} parametrization, in which every
$S$-matrix graph is built of
{\em minimal propagators} and
{\em minimal vertices}. The numerator of the minimal propagator is
just a covariant spin sum considered as a function of four independent
components of momentum%
\footnote{
The conventional way to extend the spin sum --- numerator of the
propagators --- out of the mass shell is explained, e.g., in
\cite{WeinMONO} Chap.~6.2. In principle, it can be done in many ways,
so that additional
{\em regular} terms may arise in propagator. In
\cite{AVVV2} we allowed those terms just for the sake of generality,
calling the resulting structure as ``non-minimal'' propagator.
However, as one can deduce from the discussion in the cited above
Chapter, these ``non-minimal'' items can always be cancelled by adding
certain local terms in the Hamiltonian. That is why from now on we
shall use the minimal (sometimes called transverse) propagators only.
Besides, due to peculiar features of  spin sums for massless
particles, we imply that the Hamiltonian does not contain the massless
fields of spin
$J \geq 1$. The latter is quite sufficient for work with the hadron
spectrum.
}.
The essence of the term ``minimal'' when relates to a scalar function
is that the latter looks similar on and off the mass shell, and when
relates to a tensor structure, is that it does not vanish when dotted
by relevant wave function. The central object is the
{\em minimal effective vertex}.

To explain what it is, we start from the basic Hamiltonian (without
counterterms). Let us single out all its items constructed from a
given set of, say,
$n$ normally ordered field operators with quantum numbers collectively
referred to as
$i_1,\ldots,i_n$. These items differ from each other by the
Hamiltonian coupling constants, by the number of derivatives and/or,
possibly, by their matrix structure due to fermions or a linear
symmetry group. Now, consider a momentum space matrix element of the
(formal) infinite sum of all these terms. This matrix element should
be calculated
{\em on the mass shell}, presented in a Lorentz-covariant form and
considered as a function of
$4(n-1)$ {\em independent} components of particle momenta
$p_k^\mu$ (four-momentum conservation
$\delta$-function is retained, but on-shell restriction is relaxed).
The wave functions should be crossed out. The resulting structure we
call as
{\em $n$-leg minimal effective vertex of the Hamiltonian level}. Every
such vertex
$V^{({\scriptscriptstyle \cal H})}$ presents a finite sum%
\footnote{
Cf.
\cite{AVVV2}, Eqs.~(4)--(7) and Eqs.~(12)--(13).
}
\[
V^{({\scriptscriptstyle \cal H})}_{\ldots}
(i_1,\ldots,i_n;\, p_1,\ldots,p_n) 
\]
\begin{equation}
= \delta(\Sigma p_k)\; \sum_{a} T_{\ldots}^a
V_a^{({\scriptscriptstyle \cal H})\, i_1\ldots i_n}
(\nu_1,\ldots,\nu_{3n-10})\ ,
\label{eq:h-level-vertex}
\end{equation}
of tensor/matrix structures
$T^a_{\ldots}$ dotted by scalar formfactors
$V_a^{({\scriptscriptstyle \cal H})}$ linear in Hamiltonian coupling
constants, each formfactor being a formal power series in relevant
scalar kinematical variables
$\nu_1,\ldots,\nu_{3n-10}$ (the amount of independent scalars formed
of
$p_k^\mu$ that can survive on-shell is
$3n-10$):
\[
V_a^{({\scriptscriptstyle \cal H})\, i_1\ldots i_n} (\ldots) =
\sum_{r_1,\ldots,r_d = 0}^{\infty}
g^{(a, {\scriptscriptstyle \cal H})\, i_1\ldots i_n}_{ r_1 \ldots r_d}
\;
\nu_1^{r_1} \ldots \nu_d^{r_d}\ ,
\]
\[
d = 3n - 10;
\]
here
$g^{(a, {\scriptscriptstyle \cal H})\, i_1\ldots i_n}_{
r_1 \ldots r_d}$ stand for linear combinations of the Hamiltonian
couplings.

We use the term ``minimal vertex of the Hamiltonian level'' (not
effective) to denote any separate contribution to the above series ---
a momentum space vertex produced by the basic Hamiltonian which does
not alter on-shell when dotted by the wave functions of external
particles. Except the trivial cases like
$\phi^4$, a Hamiltonian term (like, e.g.
$\phi^2\partial_\mu\phi\partial^\mu\phi$) gives rise to both minimal
and non-minimal vertices, that is why the notion of minimality makes
sense in momentum space only. It is also sensible to the choice of
variables
$\nu_k$, thereby in actual calculations one should fix this choice
which, however, is not essential here.

\begin{figure}[!th]
\includegraphics{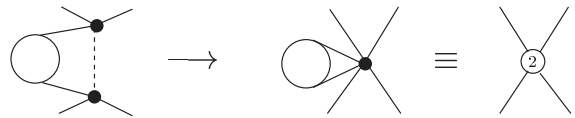}
\caption{
Example of graph reduction: the dotted propagator is cancelled by
non-minimal (vanishing on-shell, or when multiplied by the wave
function forming the propagator numerator) structure coming from one
of the vertices. Two initial vertices merge to form a secondary
vertex, 2-loop bubble-like structure is absorbed in the coupling of
new 2-nd level vertex.
\label{fig:1}
}
\end{figure}

\begin{figure*}[!th]
\begin{center}
\includegraphics{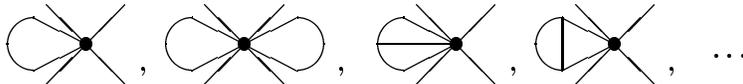}
\end{center}
\caption{
Vertices of the levels
$l=1,2$ pictured via Hamiltonian vertices: bubbles just rescale
couplings and increase the level index.
\label{fig:2} }
\end{figure*}

Minimal vertices of tree and higher levels can be constructed after
all the amplitude graphs%
\footnote{
Those computed on the mass shell of all external particles and dotted
by the relevant wave functions.
}
are subjected to the
{\em reduction procedure}
\cite{AVVV2}. In this process some of propagators disappear being
cancelled, e.g., by
$(p^2 - M^2)$ factors from non-minimal vertices. The two vertices
connected by such propagator flow together to form a single
{\em secondary vertex}, like e.g. in
Fig~\ref{fig:1}.
To preserve loop counting, we assign to this new vertex the
{\em level index} equal to the loop order of the initial structure
reduced (contracted) to form the vertex
{\em plus} the loop order of bubble-like structure%
\footnote{
But not
{\em tadpole}-like one. Note that in
\cite{AVVV2} we considered tadpoles
($1$-leg graphs) attached to a given vertex on the same footing as
self-closed lines. Here, however, we consider tadpoles as independent
elements of Feynman rules for constructing graphs in terms of
resultant parameters. This allows us to avoid (rather formal) problems
with the definition of one-particle irreducibility. Anyway, the
tadpoles can always be removed by relevant renormalization
prescription --- see
Sec.~\ref{sec-RP3}.
}
got attached to this vertex after the reduction of all graphs is
completed
(Fig.~\ref{fig:2}).
For example, if two Hamiltonian vertices  were connected with one
another by one
$(l-k)$-loop self-energy subgraph  and, in addition, by the
$k$ simple propagators, the reduction of the latter ones leads to
appearing of a new --- secondary --- vertex of the
$l$-th level: see e.g.
Fig.~\ref{fig:1}, where
$k=1$ and
$l=2$. Further, if one of the initial vertices was, say, 1-loop
counterterm, then the level assigned to the secondary vertex is
$l+1$, and so on, so that the initial loop order is kept. The idea,
of course, is that self-closed lines do not alter the tensor/matrix
structure of the vertex, only rescaling the vertex coupling (the
regularization is implied). Thus it is natural to treat the vertex
together with bubbles as a new single vertex, where a new coupling is
given by the product of the two (or more) initial ones, times whatever
bubbles give. However, since there are hidden loops, this new vertex
should not appear in calculations until the needed loop order is
reached.

In
\cite{AVVV2} we have shown that after full reduction is done, all the
amplitude graphs are expressed via the minimal propagators and minimal
vertices, some of them being the secondary vertices of various levels.
All of them are minimal in a sense that they do not change their
Lorentz-covariant form when put on mass shell and multiplied by the
relevant wave functions. All bubble-like structures disappear forming
minimal vertices of higher levels.

Now, we define the
{\em $0$-th (tree) level minimal effective vertex}
as that of the Hamiltonian level
{\em plus} the sum of all
$0$-th level secondary vertices with the same external legs. Clearly,
the most general tensor structure of the vertex is defined by the
external legs only, therefore the tensor structure of the tree level
minimal effective vertex is the same as that of the Hamiltonian level
vertex (\ref{eq:h-level-vertex}). However, due to secondary vertices,
the formfactors
$V_a^{(0)\, i_1\ldots i_n}$ are not anymore linear in Hamiltonian
couplings.

The
{\em $l$-th level} minimal effective vertex with the same set of
external legs is just a sum of all
$l$-th level minimal vertices with those legs, without adding the
{\em $l$-th level minimal effective counterterm vertex} with the same
legs%
\footnote{
Being considered separately, an
$l$-th level minimal effective counterterm vertex is built of
counterterm vertices of the
$l$-th loop order in the same way as the Hamiltonian level minimal
effective vertex is built of the Hamiltonian vertices. Of course, it
has the same tensor/matrix structure.
}.
As we just mentioned, the presence of bubbles does not change the
tensor structure of the effective vertex, it only affects the
coefficients of scalar formfactors. These latter coefficients
(eventually supplied with the index
$l$) are called the
{\em $l$-th level minimal parameters}.

Consider now a process involving a given set of external particles.
Along with other graphs, the renormalized
$l$-th loop order amplitude of this process acquires contributions
from both
$l$-th level minimal effective vertex and
$l$-th level minimal effective counterterm with the same set of
external lines. Since both vertices have the same tensor
structure, we can finally combine them into a single effective
vertex, which we call the
{\em resultant vertex} of the
$l$-th level. Simply speaking,
{\em we allow awkward vertices with bubbles and those came from the
reduction of non-minimal elements of graphs produced by the initial
Feynman rules to be absorbed by the relevant counterterms of the
corresponding loop order} and consider the resulting combination as a
single item. Analogous to minimal parameters, the
{\em $l$-th level resultant parameters} (couplings) are the
coefficients in formal power series representing relevant formfactors
or, in general, any other set of independent parameters describing the
resultant vertex. They are, of course, the functions of initial
Hamiltonian couplings. However the latter functional dependence is not
of interest anymore: we are not going to fix any of couplings in the
initial Hamiltonian. Rather, we will prefer to operate with minimal or
resultant parameters directly. The simplest case is the 3-, 2- and
eventual 1-leg resultant vertices. One can easily check, that when put
on shell and multiplied by relevant wave functions, the 1- and 2-leg
vertices do not depend on external momenta, while those with 3 legs
can only depend on it through the tensor structures like
$p^\mu$ or
$\gamma^\mu$. Hence, the formfactors in corresponding minimal
effective (and resultant) vertices are reduced to constants.

Reduction technique introduced in
\cite{AVVV2} allows one to show that any amplitude graph of loop order
$L$ can always be presented as a sum of graphs built of minimal
propagators and the minimal vertices of the levels
$l \leq L$. In turn, the full (renormalized) sum of such
$L$-th order graphs describing certain scattering process can be
re-expressed solely in terms of minimal propagators and the resultant
vertices
${V_{res}^{(l)}}_{\cdots}$ with level indices
$l \leq L$. Therefore, as long as the
$S$-matrix is considered, the only building blocks we need in the
Feynman rules are the resultant vertices and the minimal propagators.

The special convenience of dealing with the set of resultant
parameters is that it is full and its members are independent. It is
full in the sense that no other constants are needed to describe the
renormalized
$S$-matrix elements of the
$L$-th loop order but the resultant parameters with
$l \leq L$. They are independent in the sense that taking account of
the higher loop order
$l > L$ graphs leaves the structure of the lower level
$l \leq L$ parameters unchanged. The reason for this is of course a
freedom in the counterterm couplings which we consider independent at
this stage. Thus, two resultant vertices
$V^{(l_1)}$ and
$V^{(l_2)}$ with the same external lines but of different levels
$l_1$ and
$l_2$ are described by precisely the same tensor structures, but the
coefficients in power series (in the same set of variables) --- the
resultant parameters --- do not depend on each other, which is
indicated by different level indices%
\footnote{
Analogous statement was made in
\cite[p.9]{AVVV2} with respect to
{\em minimal parameters} of different levels. This is not quite
correct, until all the counterterms are taken into account and, hence,
the resultant parameters are formed.
}.
Besides, by the very construction, the resultant parameters of the
same level are independent, as far as we consider independent all the
coupling constants in the effective Hamiltonian.

However, there is one thing, unpleasant from the technical point of
view, that happens during the reduction%
\footnote{
It does not affect the tree-level calculations of
\cite{VVV}--\cite{POMI}, neither the results of
\cite{AVVV2}, thereby it was not mentioned there.
}.
As above, suppose that one works with regularized expressions. Before
the reduction one could think about all the amplitude graphs at any
loop order as being finite: the counterterms were adjusted in a way
that all subdivergencies for each given graph are cancelled when
regularization is removed. As it is clearly seen, the reduction is
nothing but rearrangement of parameters within the graphs of a given
loop order
--- it does not change the values of
$S$-matrix elements. Imagine, however, that some graph had a subgraph
with the divergency proportional to
$p^2 - M^2$, where
$p$ and $M$ are the momenta and mass of a particle on corresponding
external (w.r.t. the subgraph) leg. Before the reduction this
subdivergency had been removed by the relevant explicitly drawn
counterterm vertex of the form
\[
- C(p^2 - M^2) + \ldots\ ,
\]
where
$C$ possesses exactly the same singular behavior w.r.t. regulator as
the relevant part of the subgraph. But during the reduction the
situation changes. Due to non-minimal structure,
$(p^2 - M^2)$, the corresponding propagator disappears from the graph
and the non-minimal counter\-term gets absorbed by the new (secondary)
vertex. As a result, we may have a subgraph with divergency
proportional to
$(p^2 - M^2)$ and no
{\em explicit} counterterm to kill it. Instead, one of new couplings
acquire singular behavior so that the
$S$-matrix remains finite. This is technically inconvenient, because
one is then forced to keep working with regularization until all the
amplitude graphs of a given loop order are calculated.

Looking for remedy, one may find convenient to re-introduce some
non-minimal counterterms after the reduction is done. This, in turn,
may require renormalization prescriptions fixing the relevant
non-minimal parameters. Since, as stressed above, the
$S$-matrix does not depend on the latter, the only thing one needs to
take care of is that the chosen values of non-minimal quantities do
not fall in contradiction with various self-consistency relations. We
shall treat this technical problem in forthcoming publication. In this
paper we just assume that it is solved in one way or another (see also
the discussion in
Sec.~\ref{sec-RP2} and
Appendix~\ref{sec-WFren}).

It is now clear that renormalization prescriptions (RP's) required to
calculate the finite
$S$-matrix are of two types. The first type RP's restrict the
off-shell behavior of subgraphs. These prescriptions play no role in
fixing the on-shell value of the graph itself; they are only needed to
make convenient the intermediate steps of amplitude calculations and,
in principle, would be required to get finite Green functions. We do
not consider them in this paper. In contrast, the second set of RP's
(called below as
{\em minimal}) fixes the finite parts of counterterm constants
contained in resultant parameters, which determine the value of each
$S$-matrix element. Therefore the first step towards reducing the
required number of independent RP's is to study the structure of this
latter set.

There are certain subtleties in the usage of resultant vertices for
constructing the amplitude. First, the only possible tadpoles are the
$1$-leg resultant vertices which, as explained in
Sec.~\ref{sec-RP3}, can be safely dropped. Next, as mentioned above,
the self-closed lines are also not present anymore being absorbed in
resultant vertices. It makes no sense to picture explicitly those
bubbles, because the resultant couplings are independent parameters of
a theory. Therefore, due to ``hidden'' loops present in vertices with
levels
$l \geq 1$, the true loop order of a graph
$L_{true}$ may differ from the number of explicitly drawn loops
$L_{expl}$. To keep the right loop order, we have to take account of
the level index
$l_i$ of each resultant vertex
$V_i^{(l_i)}$. The true loop order is then given by the number of
explicitly drawn loops plus the sum of levels of the vertices used to
construct the graph under consideration:
\begin{equation}
L_{true} = L_{expl} + \sum_{vertices} l_i\ .
\label{1.2}
\end{equation}

We can now sum up the results of
\cite{AVVV2} discussed in this Section in a form of instruction. To
construct the
$L$-loop contribution to the amplitude of a given process in the
framework of effective scattering theory, one needs to:

1) Use the system of Feynman rules only containing the minimal
propagators and minimal effective (resultant) vertices of the levels
$l \leq L$.

2) Construct all the graphs with
$L_{expl} \leq L$
{\em explicit} loops with no bubbles involved and take account of
the relevant symmetry coefficients. Below
(Sec.~\ref{sec-localizability}) we argue that there is no need in
calculating amplitudes of the processes with external lines
corresponding to resonances.

3) Pick up and sum all the graphs respecting relation
(\ref{1.2}) with
$L_{true} = L$.

We will need these results in
Sec.~\ref{sec-RP1}--\ref{sec-RP3}
to explore the structure of minimal RP's needed to fix the physical
content of effective scattering theory. But first we shall formulate
two principles which we use as the basis for constructing the
well-defined finite loop order amplitudes.

\section{Basic principles of constructing the perturbation series}
\label{sec-basic}
\mbox{}

In a sense, the argumentation in this Section and in
Sec.~\ref{sec-localizability} below is inspired by the philosophy
originally developed by Krylov and Bogoliubov
\cite{Bogoliubov} for nonlinear oscillation theory. It is concerned
with perturbation series with singular behavior and allows one to
group the items in a way that the summation procedure acquires
meaning. In this spirit we specify certain requirements for the
Dyson's type series arising in the strong interaction effective
theories.

First of all one needs a parameter to put the terms of perturbation
series in certain order. Since the effective Hamiltonian involves an
infinite number of coupling constants, the conventional logic (weak
coupling or, the same, small perturbation) does not work, especially
in strong interaction physics. That is why it is commonly accepted to
classify the terms in perturbation series according to the (true)
number of loops in Feynman graphs (see, e.g.,
\cite{Coleman}).

However, in effective theories the problem of meaning of the loop
series expansion (is it convergent? asymptotic?...) is even more
intricate than in conventional renormalizable theories. Indeed, in
this case each item of the loop expansion, in turn, presents an
infinite unordered sum of graphs. This is because the interaction
Hamiltonian is a
{\em formal sum} of all possible monomials constructed from the field
operators and their derivatives of arbitrary high degree and order.
The problem of strong convergence of such operator series is not
simple, if ever meaningful in the framework of perturbation theory.
Instead, below we formulate two conditions of
{\em weak convergence} for the functional series for
$S$-matrix elements of a given loop order.

One of the most important requirements which we make use of when
constructing the meaningful items of the Dyson perturbation series is
that of polynomial boundedness. Namely, the full sum of
$S$-matrix graphs with given set of external lines and fixed number
$L$ of loops must be polynomially bounded in every pair energy at
fixed values of the other kinematical variables. There are two basic
reasons for imposing this limitation. First, from general postulates
of quantum field theory (see, e.g.,
\cite{axioms})
it follows that the full (non-perturbative) amplitude must be a
polynomially bounded function of its variables. Second, the experiment
shows that this is quite a reasonable requirement. Since we never fit
data with non-perturbative expressions for the amplitude, it is
natural to impose the polynomial boundedness requirement on a sum of
terms up to any fixed loop order and, hence, on the sum of terms of
each order. Similar argument also works with respect to the bounding
polynomial degrees. To avoid unnecessary mutual contractions between
different terms of the loop series, we attract the following
{\em asymptotic uniformity} requirement:
{\em the degree of the bounding polynomial which specifies the
asymptotics of a given loop order amplitude must be equal to that
specifying the asymptotics of the full (non-perturbative) amplitude of
the process under consideration.}
Surely, this latter degree may depend on the type of the process as
well as on the values of the variables kept fixed%
\footnote{
This is a generalization of the requirement first suggested in
\cite{WeinbARCS}, see also
\cite{ARCS}.
}.

The condition of asymptotic uniformity (or, simply, uniformity) is
concerned with the asymptotic behavior of the total contribution at
some fixed loop order, but does not tells us
{\em how} the unordered infinite sum of graphs with the same number of
loops (and, of course, describing the same process) can be converted
into the well-defined
{\em summable}%
\footnote{
This is a loan term widely used in modern theory of divergent series;
see, e.g.,
\cite{Balser}.
}
functional series. To solve the latter problem we rely upon another
general principle which we call
{\em summability requirement}%
\footnote{
By analogy with the maximal analyticity principle used in the analytic
theory of
$S$-matrix (see, e.g.
\cite{Chew}) sometimes we call it as
{\em analyticity principle}.
}.
It is formulated as follows:
{\em in every sufficiently small domain of the complex space of
kinematical variables there must exist an appropriate order of
summation of the formal series of contributions coming from the graphs
with given number of loops, such that the reorganized series
converges. Altogether, these series must define a unique analytic
function with only those singularities that are present in individual
graphs.}

At first glance, the summability requirement may seem somewhat
artificial. This is not true. There are certain mathematical and
field-theoretical reasons for taking it as the guiding principle that
provides a possibility to manage infinite formal sums of graphs in a
way allowing to avoid inconsistencies. It is, actually,
{\em both} the summability and uniformity principles that allow us to
use the Cauchy formula to obtain well defined expression for the
amplitude of a given loop order. This will be demonstrated many times
in the rest of the article.

We would like to stress that the requirements of uniformity and
summability are nothing but independent subsidiary conditions fixing
the type of perturbation scheme which we only work with. Surely, there
is no guarantee that on this way one can construct the most general
expressions for the
$S$-matrix elements in effective theory. Nevertheless, there is a
hope to construct at least meaningful ones presented by the Dyson's
type perturbation series only containing the well-defined items.

\newpage

\section{The Cauchy formula in hyper-layers}
\label{sec-Cauchy}
\mbox{}

Applying the famous Cauchy integral formula to the scattering
amplitude is a basic tool of the analytic
$S$-matrix
approach and it is very well treated in the literature. We also use
this tool but in a way essentially different from the conventional
one.

First, we apply the Cauchy formula to the
{\em finite loop order amplitudes}.
Second, being armed with the polynomial boundedness principle
discussed in the previous Section, we pay special attention to the
{\em convergency} of resulting series of integrals. Basically, we
treat the Cauchy integral as the tool to put in order the so far
unordered scope of Feynman graphs of a given loop order in a way that
the resulting series converge and, therefore, make sense. This turns
out especially useful when we need to express the amplitude of a given
loop order in terms of resultant vertices and for deriving bootstrap
equations for the physical parameters. For use in the rest of the
article and for future references we shall thus outline the main steps
of the Cauchy integral formula application.

\begin{figure}[!htb]
\includegraphics{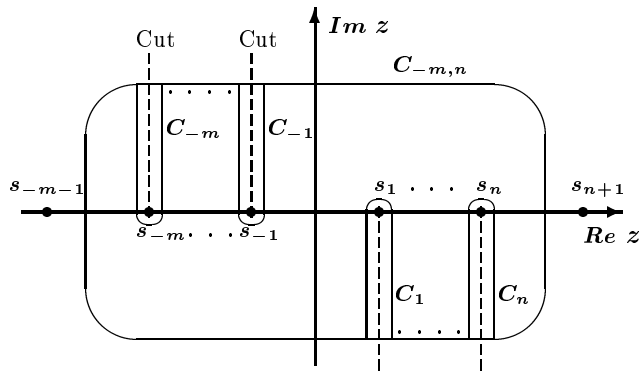}
\caption{
System of embedded contours on the complex-$z$ plane.
\label{fig:sys-of-con}
}
\end{figure}

Consider a function
$f(z,{\bf x})$ analytic in the complex variable
$z$ and smoothly depending on a set of parameters
${\bf x} \equiv  \{ x _i \}$. Suppose further that when
${\bf x} \in D$, where
$D$ is a small domain in the parameter space, this function has only a
finite number of singular points in every finite domain of the
complex-$z$ plane. In
Fig.~\ref{fig:sys-of-con} 
it is shown the geography of singular points
$s_k \equiv s_k (\bf x)$,
$k = \pm 1, \pm 2, \ldots$, typical for the finite loop order
scattering amplitudes in quantum field theory. Both left and right
singular points are enumerated in order of increasing modulo. Note
that the cuts are drawn in unconventional way --- just to simplify the
figure. If the point
$s_k$ corresponds to the pole type singularity there is no need in a
cut, but its presence makes no influence on the following discussion.

Let us recall the definition of the polynomial boundedness property
adjusted for the case of many variables
\cite{AVVV1}. Consider the system of closed embedded contours
$C(i) \equiv  C_{-m_i,n_i}$
(Fig.~\ref{fig:sys-of-con}) such that every
$C(i)$ surrounds
$C(i-1)$ and does not cross the singular points.
We say that the function
$f(z,{\bf x})$ is
$N$-bounded in the hyper-layer
$B_x\{z \in {\bf C},\, {\bf x} \in D \}$ if there is an infinite
system of contours
$C(i)$, $i=1,2,\ldots$, and an integer
$N$ such that
\begin{equation}
\max_{{\bf x} \in D;\ z \in C(i)}
{\left|\frac{f(z,{\bf x})}{z^{N+1}}\right|}
\longrightarrow 0
\label{3.1}
\end{equation}
when
$i \rightarrow \infty$. The
{\em minimal}
$N$ (possibly, negative), which provides the correctness of the
uniform (in
$x$) estimate
(\ref{3.1}), we call the degree of bounding polynomial in the layer
$B_x$.

Instead of the precise definition given above, one can just keep in
mind the rough condition, more ``strong'':
$f({\bf x},z) = o(|z|^{N+1})$ for all
${\bf x} \in D$ and large
$|z|$, except small vicinities of singularities.

Condition
(\ref{3.1}) makes it natural to apply the Cauchy's integral
formula for the function
$f(z,{\bf x})/z^{N+1}$ on the closed contour formed by
$C(i)$ (except small segments crossing the cuts), the corresponding
parts of the contours
$C_k$, $k=-m_i,...,-1,1,...,n_i$, surrounding cuts, and a small circle
around the origin%
\footnote{
We assume that
$f$ is regular at the origin. Therefore
$f(z,{\bf x})/z^{N+1}$ may have a pole there and to apply the Cauchy
formula one should add a circle around the origin to the contour of
integration. It is this part of the contour that gives the first sum
in the right side of
Eq.~(\ref{3.2}).
}
(the last one is not drawn in
Fig.~\ref{fig:sys-of-con}). In the limit
$i \to \infty$ one obtains:
\[
f(z, {\bf x})
= \sum \limits_{n=0}^{N}
\frac{1}{n!} f^{(n)}(0, {\bf x}) z^n
\]
\begin{equation}
+\ \frac{z^{N+1}}{2 \pi i}
\sum \limits_{k= -\infty}^{+\infty}\;
\oint\limits_{C_k}
\frac{f(\xi , {\bf x})}{{\xi}^{N+1} (\xi - z)}\; dz \ ,
\label{3.2}
\end{equation}
with
$f^{(n)}(z,{\bf x}) \equiv \partial^n/\partial z^n\, f(z,{\bf x})$. It
is essential to perform the last summation
{\em in order of increasing modulo of the singularities
$s_k$} which the contours
$C_k$ are drawn around. The equation
(\ref{3.2}) provides a mathematically correct form of the result. If
the number of singular points is infinite, every contour integral on
the right side should be considered as a single term of the series.
The mentioned above order of summation provides a guarantee of the
{\em uniform} (in both
$z$ and
${\bf x}$) convergence of the series. The formula
(\ref{3.2}) plays the key role in the renormalization programme
discussed below.

If the function
$f$ represents the
{\em tree level} amplitude, the summability principle formulated in
Sec.~\ref{sec-basic} does not permit any brunch cut, because only the
pole type singularities appear in tree level graphs. Hence, all the
contours
$C_k$ are reduced to circles around poles and all the integrals in
Eq.~(\ref{3.2}) can be expressed via the relevant residues. It is this
way that the
{\em Cauchy forms} introduced in
\cite{AVVV1} arise. For future reference we discuss this case in
Appendix~\ref{sec-AppCauchy}.

\section{Minimal prescriptions 1: tentative consideration}
\label{sec-RP1}
\mbox{}

The reason to construct resultant parameters shortly discussed in
Sec.~\ref{sec-prelim} is to single out the renormalization
prescriptions (RP's) needed to calculate scattering amplitudes
perturbatively. In turn, the results of
Secs.~\ref{sec-basic} and
\ref{sec-Cauchy} give a hand in forming the expressions for given loop
order amplitude in terms of resultant parameters. The following three
Sections demonstrate how all this works together in explicit
amplitude calculations. Very important result is formulated in
Sec.~\ref{sec-RP3}. Namely, it is shown that, under certain
assumptions suggested by phenomenology, in the effective scattering
theory of strong interactions one only needs to know those minimal
RP's which fix the resultant vertices with 1, 2 and 3 external legs.
The other resultant couplings turn out to be fixed by certain
self-consistency conditions. To show the origin of these conditions we
discuss below a simple example illustrating the main idea of our
renormalization procedure.

Consider an elastic scattering process
\begin{equation}
a(p_1) + b(k_1) \to a(p_2) + b(k_2)\ ,
\label{4.1}
\end{equation}
where we took both
$a$ and
$b$ particles to be spinless: this considerably simplifies the purely
technical details without changing the logical line of the analysis.

\begin{figure}[ht]
\begin{center}
\includegraphics{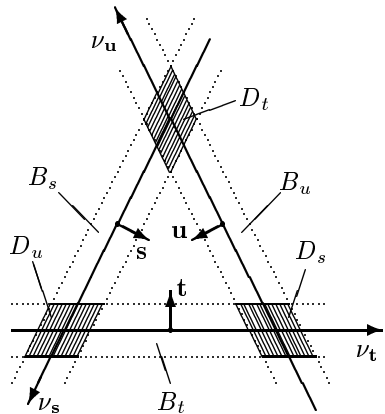}
\end{center}
\caption{Disposition of the bands
$B_x$ (bounded by dotted lines) and intersection domains
$D_x$ (hatched) $(x = s , t , u)$.
\label{fig:triangle}
}
\end{figure}

Along with the conventional kinematical variables
$s = (k_1 + p_1)^2$, $t = (k_1 - k_2)^2$ and
$u = (k_1 - p_2)^2$, we introduce three equivalent pairs of
independent ones:
\begin{equation}
(x, {\nu}_x), \quad x=s,t,u; 
\label{4.2}
\end{equation}
\[
{\rm where} \;\;
{\nu}_s = u - t,\;
{\nu}_t = s - u,\;
{\nu}_u = t - s.
\]
The pair
$(x, {\nu}_x)$ provides a natural coordinate system in 3-dimensional
(one complex and one real coordinate) hyper-layer
$B_x\{\nu_x\in {\bf C};\; x\in {\bf R};\; x\sim 0 \}$, while the pair
$(x, {\rm Re}\; {\nu}_x)$ does the same in the band parallel to the
corresponding side
$x=0$ of the Mandelstam triangle:
Fig.~\ref{fig:triangle}.

Let us suppose that in
$B_t\{\nu_t\in {\bf C};\; t\in {\bf R},\; t\sim 0 \}$
the full (non-perturbative) amplitude of the process
(\ref{4.1}) is described by the 0-bounded function
$f(\nu_t, t)$ ($N_t \leq 0$). It is quite a typical experimental
situation in hadron physics; in the end of this Section we discuss
more involved cases. According to the uniformity principle
(Sec.~\ref{sec-basic}), we have to construct the perturbation series
\[
f({\nu}_t, t) = \sum \limits_{l=0}^{\infty} f_l({\nu}_t, t)
\]
in such a way that each full sum
$f_l(...)$ of the
$l$-th loop order graphs also presents the 0-bounded function in
$B_t$. Hence, the relation
(\ref{3.2}) in this layer reads
\begin{equation}
f_l({\nu}_t,\; t ) = f_l(0,t) +
\frac{{\nu}_t}{2 \pi i}
\sum \limits_{k = -\infty}^{+\infty}\;
\oint\limits_{C_k (t)}^{}
\frac{f_l(\xi , t)}{{\xi} (\xi - {\nu}_t)}\; d{\xi}\ .
\label{4.3}
\end{equation}
Here the notation
$C_k (t)$ is used to stress that the positions of singularities (and,
hence, of the cuts) in the complex%
-${\nu}_t$ plane depend upon the other variable
$t$, which now serve as a parameter.

When working at loop order
$l$ with 4-leg amplitude we, of course, imply that all the numerical
parameters needed to fix the finite (renormalized) amplitudes of the
previous loop orders, as well as those fixing 1-, 2- and 3-leg
$l$ loops graphs are known and one only needs to carry out the
renormalization of the
$l$-th order 4-legs graphs. In the next Section we will prove that the
infinite sum of integrals in
(\ref{4.3}) depends solely on the parameters already fixed on the
previous steps of renormalization procedure. Thus to obtain the
complete renormalized expression for the
$l$-th order contribution in
$B_t$ it only remains to specify the function
$f_l(0, t)$. This can be done with the help of self-consistency
requirement.

To make use of this requirement we consider the cross-conjugated
process
\begin{equation}
a(p_1) + \overline{a}(-p_2) \to  \overline{b}(-k_1) +b(k_2)
\label{4.4}
\end{equation}
and suppose that in
$B_u\{{\nu}_u \in {\bf C};\; u \in {\bf R},\; u \sim 0 \}$
it is described by the (-1)-bounded
($N_u \leq -1$; we discuss the other possibilities below) amplitude
\[
\phi ({\nu}_u,\; u) =
\sum \limits_{l=0}^{\infty} {\phi}_l({\nu}_u,\; u)\ .
\]
The uniformity principle tells us that every function
${\phi}_l({\nu}_u,\; u)$, in turn, must be (-1)-bounded in
$B_u$ and, hence,
(\ref{3.2}) takes the form
\begin{equation} {\phi}_l({\nu}_u,\; u ) = \frac{1}{2 \pi i}
\sum\limits_{k = -\infty}^{+\infty}\;
\oint\limits_{C_k (u)}
\frac{{\phi}_l(\xi , u)}{(\xi - {\nu}_u)}\; d{\xi}\ .
\label{4.5}
\end{equation}
Again, it is implied (and proved in the next Section) that the sum of
integrals on the right side only depends on the parameters already
fixed on the previous steps of the renormalization procedure.

Recalling that both expressions
(\ref{4.3}) and
(\ref{4.5}) follow from the same infinite sum of
$l$-loop graphs (perturbative crossing symmetry!) and attracting the
summability principle, we conclude that in the intersection domain
$D_s \equiv B_t \cap B_u$ they must coincide with one another:
\[
f_l(0,t) +
\frac{\nu_t}{2 \pi i}
\sum \limits_{k = -\infty}^{+\infty}\;
\oint\limits_{C_k (t)}
\frac{f_l(\xi , t)}{\xi (\xi - \nu_t)}\; d\xi 
\]
\[
= \frac{1}{2 \pi i}
\sum \limits_{k = -\infty}^{+\infty}\;
\oint\limits_{C_k (u)}
\frac{\phi_l(\xi , u)}{(\xi - \nu_u)}\; d\xi\ ,
\]
which means that in
$D_s$
\[
f_l(0,t) = -\frac{\nu_t}{2 \pi i}
\sum\limits_{k =-\infty}^{+\infty}\;
\oint\limits_{C_k (t)}
\frac{f_l(\xi , t)}{\xi (\xi - \nu_t)}\; d\xi 
\]
\begin{equation}
+\ \frac{1}{2 \pi i}
\sum \limits_{k = -\infty}^{+\infty}\;
\oint\limits_{C_k (u)}
\frac{{\phi}_l(\xi , u)}{(\xi - \nu_u)}\; d\xi\
\equiv \Psi^{(0,-1)} (t,u)\ .
\label{4.7}
\end{equation}

The relation
(\ref{4.7}) only makes sense in
$D_s$. Given the asymptotics in
$B_s$, it is not difficult to construct two more relations of this
kind, one of them being valid in
$D_t \equiv B_u \cap B_s\,$, and the other one --- in
$D_u \equiv B_s \cap B_t\,$. These relations play a key role in our
approach because they provide us with a source of an infinite system
of bootstrap conditions. To explain what is bootstrap, let us consider
(\ref{4.7}) in more detail and make two statements.

First one:
{\em despite the fact that
(\ref{4.7}) only makes sense in
$D_s$, it allows one to express the function
$f_l(0,t)$ in the layer
$B_t$ in terms of the parameters which, by suggestion, have been fixed
on the previous steps of renormalization procedure}.
When translated to the language of Feynman rules, this means that in
our model example there is no necessity in attracting special
renormalization prescriptions fixing the finite part of the four-leg
counterterms. Instead, the relation
(\ref{4.7}) should be treated as that generating the relevant RP's
iteratively --- step by step. In what follows we call this ---
generating --- part of self-consistency equations as
{\em bootstrap conditions of the first kind}.

Second:
{\em the relation
(\ref{4.7}) strongly restricts the allowed values of the parameters
which are assumed to be fixed on the previous stages}.
To show this it is sufficient to note that
$f_l(0,t)$ only depends on the variable
$t$ while the function
${\Psi}^{(0,-1)} (t,u)$ formally depends on both variables
$t$ and
$u$. Thus we are forced to require the dependence on
$u$ to be fictitious. It is this requirement that provides us with an
additional infinite set of restrictions for the resultant parameters.
We call these restrictions as
{\em the bootstrap conditions of the second kind}.

The proof of both statements is simple. Let us choose
$t$ and
$u$ as a pair of independent variables. As we just mentioned,
$f_l(0,t)$ does not depend on the other variable
$u$. Using the definitions
(\ref{4.2}) and the fact that
$s + t + u = 2( m_a^2 + m_b^2 )$ ($m_a$ and
$m_b$ are the external particle masses), the variables
$\nu_t$ and
$\nu_u$ can be expressed via
$t$ and
$u$. Then if we just take some value of
$u$ within the bounds given by
$D_s$, say,
$u=0$, then the right side of
Eq.~(\ref{4.7})
\begin{equation}
f_l(0,t) = \Psi^{(0,-1)} (t,0)
\label{4.8}
\end{equation} will give us
$f_l(0,t)$ at
$t \sim 0$, and, therefore, everywhere in
$B_t$.

Further, since the domain
$D_s\{ t \sim 0,\; u \sim 0 \}$ contains the point
$(t=0,\; u=0)$, differentiating both sides of
Eq.~(\ref{4.7}) one obtains an infinite system of bootstrap conditions
of the second kind:
\[
\frac{{\partial}^k}{{\partial t}^k}
\frac{{\partial}^{m+1}}{{\partial u}^{m+1}}
\Psi_l^{(0,-1)}\Biggr|_{t=0,u=0} = 0, 
\]
\begin{equation}
k,m=0,1,\ldots\ .
\label{4.9}
\end{equation}
They restrict the allowed values of the parameters fixed on the
previous steps of renormalization procedure.

We see that the system of bootstrap conditions of a given loop order%
\footnote{
Eq.~(\ref{4.7})
does not generate the full system: it only mirrors the
self-consistency (crossing) requirements for the given order amplitude
in certain domains of the complex space of kinematical variables.
Another amplitudes/domains/orders can give additional constraints!
}
is naturally divided into two subsystems. Those of the first kind just
allow one to express certain resultant parameters via the lower level
parameters which, by condition, already have been expressed in terms
of the
{\em fundamental observables}
\footnote{
The external parameters appearing in renormalization prescriptions.
}
on the previous steps. In other words, they provide a possibility to
express some parameters in terms of observable quantities. This
subsystem does not restrict the admissible values of the latter
quantities.

In contrast, the bootstrap conditions of the second kind do impose
strong limitations on the allowed values of the physical (observable!)
couplings and masses%
\footnote{
With respect to resultant couplings this system turns out to be
homogeneous and the common scale factor remains undefined. With
respect to mass parameters this system is highly nonlinear but,
nevertheless, does not constrain the overall mass scale. This means
that at least two scaling parameters must be fixed by the
corresponding measurements (RP's).
}
of effective scattering theory with certain asymptotic conditions. In
fact, it provides us with the system of physical predictions which ---
at least, in principle --- can be verified experimentally.

To make our analysis complete we shall now explain the above-made
choice of the bounding polynomials degrees. Besides, in the next
Section we discuss the parameter dependence of contour integrals
appearing in
Eqs.~(\ref{4.3}) and
(\ref{4.5}).

~From experiment we know that the bounding polynomial degree
$N_{el}$ for the elastic scattering amplitude in
$B_t$ at
$t \sim 0$ does not exceed the value
$N=1$. As noted in
\cite{AVVV1},
if the system of contours appearing in the definition of polynomial
boundedness is symmetric with respect to the origin of, say, the
complex-${\nu}_t$
plane and the amplitude in question is symmetric (antisymmetric) in
$(s \leftrightarrow u)$,
the bounding polynomials possess the same evenness property as the
amplitude does. For simplicity, we have considered above this very
situation which occurs, e.g., in the pion-nucleon elastic scattering.
This explains why the term linear in
${\nu}_t$ is not present in
(\ref{4.3}).

In more general situation, when amplitude in
$B_t$ has a bounding polynomial degree
$N_t > 0 $ (while in
$B_u$, as above,
$N_u \leq -1$), the relation
(\ref{4.7}) is replaced by
\[
\sum\limits_{n=0}^{N_t} \frac{1}{n!} f_l^{(n)}(0,t) \nu_t^n
\]
\[
= -\ \frac{\nu_t^{N_t+1}}{2 \pi i}
\sum\limits_{k= -\infty }^{+ \infty}\;
\oint\limits_{C_k (t)}
\frac{f_l(\xi , t)}{\xi^{N_t+1} (\xi - \nu_t)}\; d\xi 
\]
\[
+\ \frac{1}{2 \pi i}
\sum \limits_{k = -\infty}^{+ \infty}\;
\oint\limits_{C_k (u)}
\frac{\phi_l(\xi , u)}{(\xi - \nu_u)}\; d\xi\
\]
\begin{equation}
\equiv \Psi (t,u)\ ,
\label{4.10}
\end{equation}
and the bootstrap conditions take a slightly different form as
compared to
(\ref{4.8}) and
(\ref{4.9}).
However, it is easy to see that our main conclusion remains unchanged:
in this case there is no necessity in attracting additional RP's.

The situation when in both layers
$B_t$ and
$B_u$ the amplitude has non-negative bounding polynomial degrees is
discussed in
Appendix~\ref{sec-Appasymp}. The analysis is similar, but one needs
also to attract RP's for some 4-leg vertices.

Running a bit ahead, we shall explain why the just considered example
deserves attention. The bootstrap equations analyzed in this Section
are valid in the intersection domain of two layers
$B_t$ and
$B_u$, which contain points with
$t \sim 0$ and
$u \sim 0$, respectively. Actually, the reason for this choice of
layers is explained by the existence of experimental information on
Regge intercepts. This choice is, however, also justified from the
field-theoretic point of view
\cite{AVVV2,AVVV1,POMI}. Here are the arguments. A formal way to
construct the
$(L+1)$-th order amplitude of the process
$X \longrightarrow Y$ is to close the external lines of the relevant
$L$-th order graphs corresponding to the process
$X + a(p_1) \longrightarrow Y + \overline{a}(p_2)$ with two additional
particles carrying the momenta
$p_1$ (let it be incoming) and
$p_2$ (outgoing). This means that the latter graphs should be
calculated at
$p_1 = p_2 \equiv q$, dotted by the
$a$-particle propagator and integrated over
$q$ (and then, of course, summed over all possible types of particles
$a$). To ensure the correctness of this procedure (see
\cite{axioms}), one needs to require the polynomial boundedness (in
$p_1$) of the
$L$-th order amplitude of the process
$X + a(p_1) \longrightarrow Y + \overline{a}(p_2)$ at
$t \equiv (p_1 - p_2)^2 = 0$ and, by continuity, in a small vicinity
of this value. Clearly, this argumentation applies to arbitrary graphs
with
$N \geq 4$ external lines. That is why bootstrap conditions would
arise and reduce the number of independent RP's even in the absence of
phenomenological data on asymptotic behavior.

\section{Minimal prescriptions 2: contribution of
         singularities}
\label{sec-RP2}
\mbox{}

As promised, here we show that all the contour integrals in
(\ref{4.3}) and
(\ref{4.5}) only depend on the parameters already fixed on the
previous stages of the renormalization procedure. The proof is based
on the structure of
Eq.~(\ref{3.2}) and on the results of
\cite{AVVV2} briefly reviewed in
Sec.~\ref{sec-prelim}. Again, we consider first the elastic two body
scattering
(\ref{4.1}), and then generalize the result to arbitrary scattering
process.

Summability principle tells us that only parameters of graphs with
singularities can appear in the contour integrals under consideration.
Working with resultant parameters, the simplest way to trace which of
them contribute to singular graphs is to picture the
$l$-th order amplitude via resultant Feynman rules --- the recipe is
given in the end of
Sec.~\ref{sec-prelim}. However, we find it instructive to demonstrate
once more how the resultant vertices are built of the Hamiltonian
ones: for this we shall look at the reduction procedure in action.
This procedure does not change the structure of singularities of a
given graph; it only re-expresses this graph via minimal parameters of
various levels. When applied to a full sum of graphs forming a given
order amplitude, it re-expresses this amplitude in terms of resultant
vertices and minimal propagators. One of great advantages of minimal
(resultant) parametrization, is that the singularities are explicitly
seen when graphs are drawn --- there are no more non-minimal
structures that could cancel the propagator's denominator.

\begin{figure*}[!th]
\includegraphics{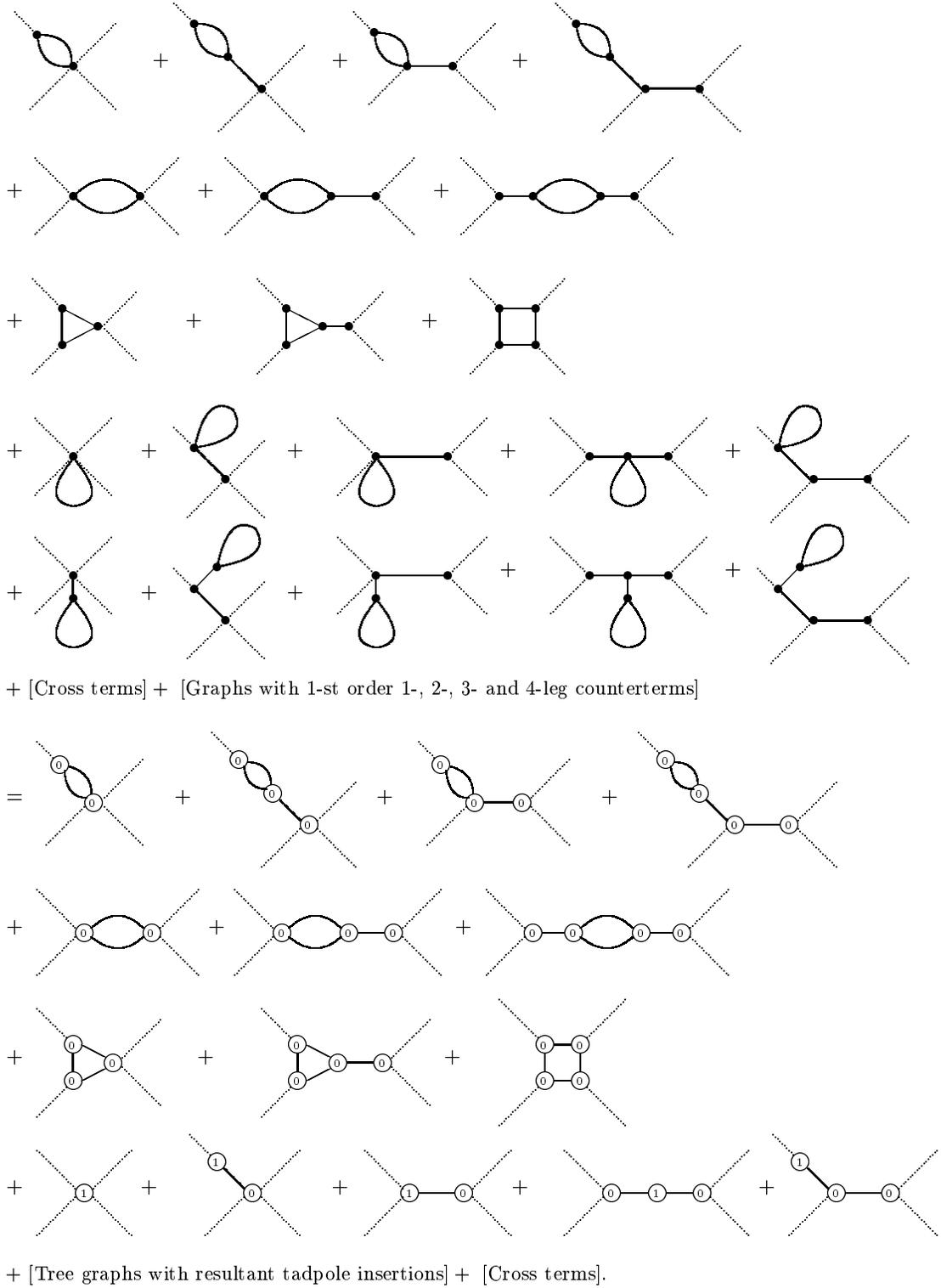}
\caption{
Reduction of
$1$-loop contribution to the
$2 \rightarrow 2$ amplitude: numbers indicate loop levels of the
resultant vertices.
\label{fig:reduct1l}
}
\end{figure*}

Let us look at one loop contribution to the process
(\ref{4.1}).
Fig.~\ref{fig:reduct1l}
schematically pictures the reduction procedure. The graphs on the left
side of pictorial equation are drawn via initial Hamiltonian
(effective) vertices%
\footnote{
Hamiltonian effective vertex (not ``minimal'') is just the sum of all
bare Hamiltonian vertices (both minimal and non-minimal) with a given
set of legs
\cite{AVVV2}. Summation over all possible internal lines and
vertices is implied in both sides of the equation.
},
that is why the self-closed lines appear. The right side is the
result of the reduction procedure: the 1-loop contribution is
presented in terms of resultant vertices of various levels (to save
space we do not draw the graphs with resultant tadpoles). The numbers
inside circles stand for the level indices of resultant vertices. One
can easily verify that
Eq.~(\ref{1.2}) and the drawing rules formulated thereafter are
respected. Namely, we see that the resulting graphs are constructed
from the minimal propagators and resultant 1-, 2-, 3- and 4-leg
vertices of levels 0 and 1. The graphs with
{\em explicit loops} only contain the vertices of the lowest level
$l=0$, otherwise the loop counting would be violated. The 1-loop
level resultant parameters appear only in the diagrams without
explicit loops: these parameters come from vertices with self-closed
lines and from 1-loop counterterms, thus no more loops are allowed.

By the  very logic of renormalization procedure, at this step the 1-,
2- and 3-leg one-loop counterterms were already adjusted to remove
infinities from the corresponding subgraphs. Thus 1-, 2- and 3-leg
resultant vertices are fixed and there are no more subdivergencies%
\footnote{
Non-minimal renormalization prescriptions may be needed to remove them
--- see the next paragraph.
}.
The only parameters which remain free are those describing 4-leg
1-loop level resultant vertex. It is this vertex that absorbs the
4-leg 1-loop counterterms, and it is the only one which remains to be
fixed by renormalization prescriptions. But this latter vertex appear
in the graph with no singular structure (contact vertex!) and thereby
cannot contribute to contour integrals around cuts (or poles) in
(\ref{4.3}) and
(\ref{4.5}).

In the end of
Sec.~\ref{sec-prelim} we already said that in this paper we only
consider the structure of
{\em minimal} RP's needed to fix the finite parts of minimal
counterterms. As shown in
\cite{AVVV2}, fixing the latter counterterms completely determines the
$S$-matrix at a given loop order. From the technical point of view it
is clear that those RP's are quite sufficient to perform the very last
step of renormalization of
$S$-matrix elements at a given loop order:
$S$-matrix is calculated on-shell and thus can be fixed by the minimal
counterterms. However, the standard way one renormalizes a graph
implies that divergencies from (off-shell!) subgraphs are removed
first. Of course, the latter divergencies are not necessary minimal
even for the subgraph built of minimal elements, and therefore may
require non-minimal counterterms%
\footnote{
For example, consider two-leg off-shell graph in
$\phi^4$ theory
($\phi^4$ is a minimal vertex, since it does not change its structure
on-shell), which has
$\Lambda^2 + p^2\log\Lambda$ behavior, so that the non-minimal ---
proportional to
$p^2$ --- counterterm is needed.
}.
The source of this apparent confusion was pointed out in
Sec.~\ref{sec-prelim} --- the non-minimal counterterms were absorbed
by minimal vertices during the reduction. Although all off-shell
subdivergencies will cancel in the given loop order amplitude, we
still have no explicit counterterm to kill each of them directly in
graphs where they arise. In particular, this relates to the wave
function renormalization. Keeping in mind what was said about possible
solutions in
Sec.~\ref{sec-prelim}, we postpone the detailed discussion to the
next publication (see, however,
Appendix~\ref{sec-WFren}). Here we will just tacitly imply that all
subgraphs are made finite. The present analysis is quite sufficient to
justify the comparison of tree-level computations with experimental
data (for preliminary discussion see
\cite{AVVV1,MENU,Talks}).

In general, the renormalization of the
$l$-loop amplitude of a process involving
$n=4,5, \ldots$ particles consists of
$l$ stages. In turn, every
$l'$-th stage
$(l'=1,2,\ldots ,l-1)$ consists of certain (depending on
$n,l$ and
$l'$) number of steps, each one being the renormalization of
$l'$-loop graphs with a given number
$1,2,\ldots , n'_{max}(n,l,l')$ of external lines. The last ---
$l$-th --- stage  consists of
$(n-1)$ preliminary steps: renormalization of the
$l$-loop resultant graphs with
$1,2, \ldots ,(n-1)$ legs (tadpole, self-energy, etc). At last, the
final
($n$-th) step --- renormalization of
$n$-leg
$l$-loop graphs --- requires attracting renormalization prescriptions
that fix the values of
$l$-th order
$n$-leg counterterm vertices. This is precisely the situation known
from conventional renormalization theory. The only difference is that
in effective scattering theory the number of counterterms needed for
renormalization of all
$l$-th loop order
$S$-matrix elements is infinite.

As stressed in the previous Section, when writing down the
$l$-th order amplitudes
(\ref{4.3}) and
(\ref{4.5}), we imply that all the previous steps already have been
passed and we only need to make the last step --- to fix the
$l$-th level 4-leg counterterms or, equivalently, fix the coefficients
in the formal series for 4-leg resultant vertex. To have a singularity
and, thus, to contribute to contour integral in
(\ref{4.3}) or
(\ref{4.5}), a graph must have at least one internal line. Using the
pictorial rules formulated in
Sec.~\ref{sec-prelim}, or just by direct analogy with
Fig.~\ref{fig:reduct1l}, it is easy to understand that graphs with
internal lines may only depend on resultant parameters of lower
levels%
\footnote{
It could be wrong if there were 2-leg resultant vertices of
$0$-th (tree) level. Or, the same, if we chose masses in propagators
to differ from the corresponding pole positions in tree-level
amplitudes. We do not --- see the next Section.
},
or on the
$l$-th level parameters from vertices with less number of legs:
$n=1,2,3$. Since all those parameters have already been fixed on the
previous steps of renormalization procedure, the contour integrals in
(\ref{4.3}) and
(\ref{4.5}) should be, indeed, considered as known functions. This
completes our proof.

Now we are in a position to review our analysis of the process
(\ref{4.1}) and put all the steps in logical order. That is, we
started from the formal sum of
$l$-loop 4-leg amplitude graphs and rewrote it in terms of resultant
parameters of various levels. We suggested that all subgraphs are
renormalized (finite) and, hence, all the resultant parameters of
lower levels
$l' = 1, 2, \ldots , l-1$ are fixed. Then we
{\em required} that this formal sum results in a function with the
following properties:

$\bullet$
In
$B_t$ it is 0-bounded in complex variable
${\nu}_t \equiv (s-u)$, while
$t$ is treated as a parameter.

$\bullet$
In
$B_u$ it is (-1)-bounded (decreasing) in complex variable
${\nu}_u \equiv (t-s)$, while
$u$ is treated as a parameter.

$\bullet$
In both layers the resulting function has only those singularities
which are presented in the contributions of individual graphs of the
formal sum under consideration (summability principle).

These requirements allowed us to rewrite the sum of graphs in a form
of Cauchy-type integral
(\ref{3.2}) which, therefore, provides the mathematically correct
expression for the
$S$-matrix element as a function of the external parameters (or, the
same, RP's) of the theory. 

It is shown that the last
($l$-th) stage of renormalization of
$S$-matrix elements does not require attracting any minimal RP's in
addition to those fixing the
$l$-th order resultant vertices with
$n=1,2,3$ legs;
{\em the RP's fixing minimal 4-leg counterterms of the
$l$-th order are automatically generated by the first kind bootstrap
equations}. Besides, the bootstrap equations of the second kind impose
certain restrictions on the
$l$-th level parameters of 1-, 2- and 3-leg vertices (and, possibly,
on the parameters of lower levels). As noted in the end of
Sec.~\ref{sec-RP1}, this conclusion actually holds for any
$2\rightarrow 2$ amplitude which possesses negative bounding
polynomial degree in one of the intersecting layers, while the
asymptotic in the other layer may be arbitrary, though also
polynomial.

This result offers a hint on which requirements are sufficient for
$S$-matrix renormalization in effective scattering theory.

\section{Minimal renormalization prescriptions: general outline}
\label{sec-RP3}
\mbox{}

The model example considered in two previous Sections shows that, as
long as there are two intersecting hyper-layers such that in both of
them the amplitude of a given process
$2 \rightarrow 2$ is polynomially bounded, and at least in one of
them it is
$(-1)$-bounded in the corresponding complex variable, there is no need
in independent RP's for 4-leg amplitude graphs: the summability
principle provides us with a tool for generating those (on-shell!)
prescriptions order by order. This conclusion, however, implies, that
all the previous renormalization steps are done, so that all subgraphs
of previous loop orders
$l' < l$ and 1-, 2- and 3-leg graphs of the same loop order
$l$ are made finite. As to the 1-, 2- and 3-leg minimal counterterms
(which are just constants in terms of the resultant parameters), one
does need to fix them by relevant RP's, but this cannot be done
arbitrarily --- the bootstrap requirements must be taken into account.

It is well known that the amplitudes of inelastic processes involving
$n > 4$ particles decrease with energy, at least in the physical area
of other relevant variables: the phase space volume grows too fast to
maintain unitarity of
$S$-matrix with non-decreasing inelastic amplitudes. Therefore it
looks natural to suggest that in corresponding hyper-layers these
amplitudes can be described with the help of at most
$(-1)$-bounded functions of one complex energy-like variable (and
several parameters). Also, it is always possible to choose the
variables in such a way that the domains of mutual intersection of
every two hyper-layers are non-empty%
\footnote{
This follows from the fact that the number of pair energies is much
larger then that of independent kinematic variables.
}.
One then easily adjusts the analysis of
Secs.~\ref{sec-RP1}--\ref{sec-RP2} to show that the amplitudes of
processes with
$n > 4$ particles are completely defined by contour integrals similar
to
(\ref{4.5}), which depend only on the parameters already fixed on the
previous steps of renormalization. In other words, RP's for 1-, 2- ,
3- and 4-leg graphs completely specify those for graphs with 5 legs,
altogether these RP's give prescriptions for 6-leg graphs, and so on.
Hence, the independent RP's for amplitude graphs with
$n > 4$ legs are not required, so the system of (minimal)
renormalization prescriptions needed to fix the physically interesting
effective scattering theory only contains prescriptions for 1-, 2-, 3-
and, possibly, 4-leg resultant graphs.

The next step is to employ hadronic phenomenology. Consider first the
$SU_2$ sector, where the only stable particles are the pion and the
nucleon. As known, the high-energy behavior of elastic pion-pion,
pion-nucleon and nucleon-nucleon scattering amplitudes is governed by
the Regge asymptotic law. Then it is not difficult to check that each
of these amplitudes is described by scalar formfactors with
{\em negative} degrees of bounding polynomial, at least in one of
three cross-conjugated channels. In fact, even for heavier flavors, we
are not aware of the process with four stable (w.r.t. strong
interactions) particles that violates this rule. That is why the
analysis in
Sec.~\ref{sec-RP1} is relevant and 4-leg amplitude graphs with stable
particles on external lines do not require formulating RP's%
\footnote{
If a process with four stable hadrons with the amplitude asymptotics
violating the Regge law will be found, additional RP's discussed in
Appendix~\ref{sec-Appasymp} may be needed.
}.

There are also 4-leg graphs with resonances on external legs. Here the
situation looks more complicated owing to the absence of direct
experimental information on processes with unstable hadrons. In other
words, the choice of relevant bounding polynomial degrees is to a
large measure nothing but a matter of postulate. The only way to check
the correctness of the choice is to construct the corresponding
bootstrap relations and compare them with existing data on resonance
parameters. This work is in progress now. Here, however, we are
tempted to consider the relatively simple situation when all the 4-leg
``amplitudes'' of the processes involving unstable particles decrease
with energy, at least at sufficiently small values of the momentum
transfer%
\footnote{
Surely, this is just a model suggestion which, however, seems us quite
reasonable. One of the arguments in its favor is that unstable
particles cannot appear in true asymptotic states
(Sec.~\ref{sec-localizability}).
}.
Then, again, according to
Sec.~\ref{sec-RP1}, the RP's for these 4-leg (on shell) graphs are not
needed.

To summarize:
{\em the only minimal RP's needed to specify all
$S$-matrix elements of the effective hadron scattering theory are
those fixing finite parts of the resultant 1-, 2- and 3-leg vertices.
Moreover, this system cannot be taken arbitrary --- the relevant
bootstrap constraints must be taken into account.} Once these basic
RP's are imposed, the minimal prescriptions for 4-, 5-, ...-leg graphs
are automatically generated at any given loop order by summability
principle. The possibility to do this for 4-leg graphs is provided by
phenomenology, for 5-, ...-leg ones --- by perturbative unitarity.
Remember, however, that some non-minimal RP's (or another way to
remove non-minimal divergencies) may also be needed.

To write the required minimal RP's explicitly we need to introduce
some notations. Consider the infinite sum
${\cal C}$ of all
$n$-leg counterterm vertices (with a fixed set of legs) of the loop
order
$l$ --- we call it the
$l$-th order effective counterterm vertex. This vertex is point-like
but
{\em not} minimal and should not be mixed with the
$l$-th level resultant vertex. We can write it as
\begin{widetext}
\begin{equation}
{\cal C}^{(l)}_{\ldots} (i_1,\ldots,i_n;\, p_1,\ldots,p_n) 
= \delta(\Sigma p_k)\;
\sum_{a}^{M+N} T_{\ldots}^a
{\cal C}_a^{(l)\, i_1 \ldots i_n}
(\pi_1,\ldots,\pi_n;\, \nu_1,\ldots,\nu_{3n - 10})\ ,
\label{7.01}
\end{equation}
were
$i_k$ marks the species of the
$k$-th particle (mass
$M_k$, spin, etc) and
$p_k$ is its four-momentum. The index
$a$ numbers tensor/matrix structures
$T_{\ldots}^a$ and ellipses stand for corresponding tensor/matrix
indices (if needed). The structures with
$a=1,...,M$ are minimal, while those with
$a=M+1,...,N$ are non-minimal --- they vanish on the mass shell when
dotted by the wave function of relevant particle. Since we work in
effective theory, all such structures allowed by Lorentz invariance
(and eventual linear symmetry) are present and do not depend on the
loop order in question. The scalar functions
${\cal C}_a^{(l) \,i_1 \ldots i_n}$ stand for the formal power series
\begin{equation}
{\cal C}_a^{(l)\, i_1 \ldots i_n}( \ldots ) 
= \sum_{s_1,\ldots, s_n,\, r_1, \ldots, r_d = 0}^{\infty}
{\cal C}^{(a, l)\, i_1\ldots i_n}_{ r_1\ldots r_d;\, s_1\ldots s_n}\;
{\nu}_1^{r_1} \ldots {\nu}_d^{r_d}
{\pi}_1^{s_1} \ldots {\pi}_n^{s_n}\ ,
\quad
d \equiv 3n - 10\ ,
\label{7.02}
\end{equation}
in ``mass variables''
$\pi_k \equiv p_k^2 - M_k^2$ ($k = 1,2,...,n$)
and ``on-shell variables''
$\nu_1, \ldots ,\nu_{3n-10}$, the latter ones are the scalar functions
of 4-momenta chosen in a way that they provide a coordinate system on
the mass shell; the concrete choice is not essential for the current
discussion.

Next, using the same notations as in
(\ref{7.01}) and (\ref{7.02}), we can present the full sum
${\cal G}^{(l)}$ of (amputated) resultant
$l$-loop%
\footnote{
Recall that the true number of loops should be calculated in
accordance with
(\ref{1.2}).
}
graphs with
$n$ external (off-shell) particles of the types
$i_1, \ldots i_n$ as follows:
\begin{equation}
{\cal G}^{(l)}_{\ldots} (i_1,\ldots,i_n;\, p_1,\ldots,p_n) =
\delta(\Sigma p_k)\;
\sum_{a=1}^{M+N}  T_{...}^a
{\cal G}_a^{(l)\, i_1...i_n}
(\pi_1,\ldots,\pi_n;\, \nu_1,\ldots,\nu_{3n-10})\ .
\label{7.05}
\end{equation}
Here
${\cal G}_a^{(l) i_1\ldots i_n}$ stand for true
$l$-th loop order formfactors which, in contrast to
${\cal C}_a^{(l) \, i_1\ldots i_n}$, are
{\em not} just formal series but complex functions of
$\pi$'s and
$\nu$'s.

Suppose that we are going to perform the
{\em last} step of renormalization --- to fix the
$l$-th loop order
$S$-matrix element for the given
$n$-particle process, while the computation of higher order terms is
not assumed. Hence, we only need RP's for the
{\em on-shell}
($\pi_k = 0$) value of that part of the sum
(\ref{7.05}) which survives when the external legs are multiplied by
the relevant wave functions. Therefore only the minimal tensor
structures contribute:
\[
{\cal G}^{(l)}_{\ldots}
(i_1, \ldots, i_n;\, p_1, \ldots, p_n)
\Bigr|_{
\stackrel{\scriptstyle\rm relevant
}{
\scriptstyle {\rm for} \; S-{\rm matrix}}
}
\]
\begin{equation}
= \delta(\Sigma p_k)\;
\prod\limits_{k=1}^n (Z_{i_k})^{\scriptscriptstyle -\frac{1}{2}}\;
\sum_{a=1}^{M}  T_{...}^a
G_a^{(l)\, i_1...i_n}(\nu_1, \ldots, \nu_{3n-10})\ ,
\label{eq:7.05a}
\end{equation}
where for
$a = 1, \ldots, M$ we have introduced
\end{widetext}
\begin{equation}
G_a^{(l)\, i_1...i_n}
\equiv
\prod\limits_{k=1}^n (Z_{i_k})^{\scriptscriptstyle \frac{1}{2}}\;
{\cal G}_a^{(l)\, i_1...i_n}  \Bigr|_{\pi_i = 0}\ ,
\label{eq:7.05b}
\end{equation}
and
$Z_{i_k}$ stands for corresponding field-strength renormalization
constant.

It is the prescriptions for
$G_a^{(l) i_1...i_n}$ that we call minimal. As it was stressed several
times in the text, fixing the latter quantities is sufficient for
specification of all the
$S$-matrix elements, though some non-minimal (off-shell) RP's may be
needed to remove infinities form subgraphs. In particular, the RP's
for derivatives of two-point functions --- field-strength
renormalization --- are exactly of non-minimal type. However, the
analysis of possible structure of non-minimal prescriptions is beyond
the scope of this article and postponed till following publications.
Here we just assume that all subdivergencies are removed from
${\cal G}^{(l)}$ and
$Z$'s are known.

The discussion in
Sec.~\ref{sec-RP2} shows that the sum
(\ref{7.05}) consists of:

{\bf a.} Graphs (not point-like) built of the lower level resultant
vertices and of resultant vertices of the
$l$-th level with lower number of legs. Couplings at these vertices
are considered fixed on the previous steps of renormalization. As we
just mentioned, all subdivergencies are removed, thereby these graphs
introduce only superficially divergent terms to be eliminated and
properly normalized by corresponding RP's.

{\bf b.} The
$l$-th level
$n$-leg resultant vertex
$V^{(l)}$ with the same set of external particles. By definition
(Sec.~\ref{sec-prelim}), the latter has the same tensor structure as
the sum of graphs
(\ref{eq:7.05a}). This vertex consists of the
$l$-th level point-like effective vertex formed after the reduction,
and the minimal (surviving on the mass shell) part of effective
counterterm vertex
(\ref{7.01}). Hence, this resultant vertex can be written as
\[
V^{(l)}_{\ldots}(i_1,\ldots,i_n;\, p_1,\ldots,p_n) 
\]
\begin{equation}
= \delta(\Sigma p_k)\;
\sum_{a=1}^M  T_{\ldots}^a
V_a^{(l)\, i_1\ldots i_n}(\nu_1,\ldots,\nu_{3n-10})\ ,
\label{7.03}
\end{equation}
where
$V_a^{(l)}$ are just formal power series:
\[
V_a^{(l)\, i_1 \ldots i_n} (\ldots) 
= \sum_{r_1, \ldots, r_d = 0}^{\infty}
g^{(a, l)\, i_1 \ldots i_n}_{ r_1 \ldots r_d} \;
{\nu}_1^{r_1} \ldots {\nu}_d^{r_d}\ ,
\]
\begin{equation}
d = 3n - 10\ ,
\label{7.04}
\end{equation}
and the numerical coefficients
$g_{ r_1 \ldots r_{3n - 10}}^{(a, l)\, i_1 \ldots i_n}$ (the
$l$-th level
$n$-leg resultant parameters) are combined from the
$l$-th level
$n$-leg coupling constants (given by products of initial Hamiltonian
couplings and bubble factors) and the corresponding
(surviving on-shell!) counterterm couplings:
\[
g_{r_1\ldots r_{3n-10}}^{(a, l)\, i_1 \ldots i_n}
= \left(
{\stackrel{
\scriptstyle\rm coupling\; of\; (secondary)
}{
\scriptstyle{\rm vertex\; with\;} l\; {\rm bubbles}}
}
\right)
+ {\cal C}^{(a,l)\, i_1\ldots i_n}_{r_1\ldots r_{3n-10};\, 0\ldots 0}
\ .
\]
To specify the
$S$-matrix contribution
(\ref{eq:7.05a}) one needs only to fix the latter sum --- as we just
mentioned, all other parameters are considered known (in fact, we just
rely on the mathematical induction method). Note however, that the
{\em off-shell} sum of graphs (the Green function) will
{\em not} be completely renormalized in this way: in general, to
compensate the off-shell superficially divergent terms one will need
to attract all the off-shell
($\pi_k \neq 0$) counterterms in
(\ref{7.01}), which may be absorbed later by the higher level
resultant parameters during the reduction of
$(l+1)$ order graphs. It is exactly the subtlety with non-minimal RP's
that we are not going to discuss further in this paper.

So, apart from the field-strength renormalization and other
non-minimal RP's, the
$S$-matrix is renormalized by adjusting the (finite part of) resultant
couplings
$g$'s in
(\ref{7.04}), which is equivalent to imposing RP's on
$G_a^{(l)\, i_1...i_n}$ in
(\ref{eq:7.05a}).

In the framework of renormalized perturbation theory one has to
identify the tree level (physical) parameters: mass parameters and
coupling constants. We define the
{\em mass parameter} (or, simply, mass)
$M$ of a particle as the number that fixes the pole position of (free)
Feynman propagator. For stable particle (like pion or nucleon) this
number must coincide with the mass of corresponding asymptotic state
or, the same, with eigenvalue of the momentum squared operator.
Resonance masses do not have such interpretation, and in
Sec.~\ref{sec-localizability} we explain how they arise. Their values
should be deduced from fit with experimental data and may depend on
various conventions (see the discussion in
Sec.~\ref{sec-resonances}).

In terms of resultant parameters, the
{\em physical coupling constants} are naturally identified with
tree-level resultant couplings
\begin{equation}
g^{(a, 0)\, i_1 \ldots i_n}_{ r_1 \ldots r_{3n-10}}
\label{eq:phys-coupling}
\end{equation}
from
(\ref{7.04}). For
$n=1,2,3$ there are no lower indices, because, as mentioned in
Sec.~\ref{sec-prelim}, the corresponding resultant formfactors are
just constants. Corresponding true formfactors
${\cal G}_a^{(l)\, i}$, ${\cal G}_a^{(l) \, i_1 i_2}$,
${\cal G}_a^{(l)\, i_1 i_2 i_3}$ are, of course,
{\em not} constants off-shell, but with our choice of variables can
only depend on
$\pi_k \equiv p_k^2 - M_k^2$.

In the first part of this Section we have shown, that the only minimal
RP's needed (at least in
$SU_2$ sector) are those fixing the values of the
resultant vertices with 1, 2 and 3 external particles. Conventionally,
renormalization prescription is imposed on the sum of graphs with a
given set of external legs computed at certain kinematical point.
Typically, the sum of all 1-particle irreducible
(1PI) graphs%
\footnote{
A reducible (in conventional sense) graph constructed from the initial
Hamiltonian vertices may become 1PI after the reduction and switching
to minimal parametrization
(Sec.~\ref{sec-prelim}): some propagator denominators may be cancelled
so that corresponding lines disappear. When working with resultant
(minimal) parameters we, of course, assume the reduction done, thereby
no confusion may arise. It is also possible to formulate the RP's for
one-particle reducible graphs (so-called
{\em over-subtractions}, see
\cite{Renorm}); here we do not consider this possibility.
}
up to a given loop order is taken
\cite{Vasiliev2, Renorm}. Equivalently, one may impose RP's on the sum
of 1PI graphs of a fixed loop order, as we do. Namely, below we imply
that the functions
${\cal G}_a^{(l)\, i_1 \ldots i_n}$ in
Eq.~(\ref{eq:7.05b}) only acquire contributions from 1PI
$l$-loop graphs, and, therefore, the
$Z$-factors are dropped. Keeping in mind that the momentum
conservation delta function
$\delta(\Sigma p_k)$ stands as an overall factor in
Eq.~(\ref{7.05}), we can now write down the required system of minimal
RP's
($l = 0,1,\ldots$):

\begin{widetext}
$\bullet$
For
$n = 1$ (absence of tadpole contributions):
\begin{equation}
{\cal G}_a^{(l)\, i_1} (p_1)
\Bigr|^{\scriptscriptstyle \rm (1PI)}_{p_1^2 = M_1^2}
\equiv G_a^{(l)\, i_1}\Bigr|^{\scriptscriptstyle \rm (1PI)} = 0\ .
\label{eq:RP-tadpole}
\end{equation}
In particular, at
$l=0$ it reads as
$g^{(a, 0) \, i_1} = 0$ so that there are no tadpoles at tree level.

$\bullet$
For
$n = 2$ (absence of mixing and real mass shift):
\begin{equation}
\widetilde{\rm Re}\; {\cal G}_a^{(l)\, i_1 i_2}(p_1, p_2)
\Bigr|_{p_k^2  = M_k^2}^{\scriptscriptstyle \rm (1PI)}
\equiv  \widetilde{\rm Re}\;
G_a^{(l)\, i_1 i_2}\Bigr|^{\scriptscriptstyle \rm (1PI)}  = 0\ .
\label{eq:RP-mass}
\end{equation}
At
$l=0$ it reads as
$g^{(a, 0) \, i_1 i_2} = 0$ so that at tree level there is neither
mixing nor correction to the particle mass.

$\bullet$
For
$n = 3$ at
$l=1,2,\ldots$:
\begin{equation}
\widetilde{\rm Re}\; {\cal G}_a^{(l)\, i_1 i_2 i_3} (p_1, p_2, p_3)
\Bigr|_{p_k^2 = M_k^2}^{\scriptscriptstyle \rm (1PI)}
\equiv \widetilde{\rm Re}\;
G_a^{(l)\, i_1 i_2 i_3}\Bigr|^{\scriptscriptstyle \rm (1PI)} = 0\ ,
\label{eq:RP-triple}
\end{equation}
while at tree level
($l=0$) this formfactor is, of course, equal to the triple physical
coupling:
\[
{\cal G}_a^{(0)\, i_1 i_2 i_3} (p_1, p_2, p_3)
\Bigr|_{p_k^2 = M_k^2}^{\scriptscriptstyle \rm (1PI)}
\equiv
G_a^{(0)\, i_1 i_2 i_3}\Bigr|^{\scriptscriptstyle \rm (1PI)}
= g^{(a,0)\, i_1 i_2 i_3}\ ,
\]
the latter we define to be real, thus attributing eventual complex
phases to the tensor structures.
\end{widetext}

Eqs.~(\ref{eq:RP-mass})--(\ref{eq:RP-triple}) adjust mass shifts
and three leg amplitudes. In fact, one is only allowed to constrain
the real parts of corresponding loop integrals, which is indicated by
the
$\widetilde{\rm Re}$ symbol. The triple couplings
$g^{(a,0) \, i_1 i_2 i_3}$ and masses
$M$ appearing in
(\ref{eq:RP-tadpole})--(\ref{eq:RP-triple}) are the
{\em physical observables}. In general, their values must be fitted
by comparison with experimentally measured amplitudes.

These prescriptions are sufficient to perform the last step of the
$S$-matrix renormalization under the above-specified conditions, of
which the most important one is the Regge-like asymptotic behavior. In
that situation when some phenomenological 4-leg amplitude has no
hyper-layers with decreasing asymptotics (with negative value of
bounding polynomial degree), one should also add prescriptions for
4-point amplitudes written down in
Appendix~\ref{sec-Appasymp}. However, as far as we know, for any
process involving four stable (w.r.t. strong interactions) particles
such hyper-layers are always present so that additional prescriptions
are not needed.

The above RP's have to be discussed. As to the tadpoles
(\ref{eq:RP-tadpole}), in the resultant parametrization they can
have only scalar particle on the external leg: the covariant
structures of type
$\partial_\mu \ldots \phi^{\mu\ldots}$ built of tensor field
$\phi^{\mu\ldots}$ do not survive on shell and thus do not
contribute to resultant parameters. So, the ``structure'' index
$a$ in
(\ref{eq:RP-tadpole}) can be omitted. The remaining scalar resultant
tadpoles give just constant factor to the graph they appear in and can
be absorbed by the relevant coupling constants. In fact, this way we
followed in
\cite{AVVV2}. Here, instead, to avoid the formal problems with
one-particle irreducibility, we just accept the
Eq.~(\ref{eq:RP-tadpole}). For calculations of amplitudes both ways are
equivalent and in the future tadpoles are dropped.

Next,
Eq.~(\ref{eq:RP-mass}) provides a definition of what we call the mass
parameter --- mass, appearing in Feynman propagator. If, as suggested,
the tadpoles are absent,
Eq.~(\ref{eq:RP-mass}) forbids any two-leg resultant vertex at 0-th
(tree) level. In the case of stable particles this looks
natural because their mass terms are attributed to the free
Hamiltonian. As to the resonances, the situation is not so
transparent; we discuss it in the next two Sections. Note also, that
the only tensor structure surviving in 2-leg resultant vertex is the
(symmetrized product of) metric tensor
$g^{\mu\nu}$ (or unit matrix for fermions) and metric tensor for
eventual linear symmetry group, therefore the ``structure'' index
$a$ in
(\ref{eq:RP-mass}) takes the only value and can be dropped. If there
are many particles with the same quantum numbers except masses, one
needs also to check that it is possible to apply RP's avoiding mixing.
This is non-trivial; we shall discuss it in the next publication.

At last, the prescription
(\ref{eq:RP-triple}) guarantees the absence of loop corrections to
the physical (real!) triple coupling constants.

Looking now at the system
(\ref{eq:RP-tadpole})--(\ref{eq:RP-triple}) and recalling the
step-stage description of the renormalization procedure given in
Sec.~\ref{sec-RP2}, it is easy to understand that at the very first
step --- calculation of tree level amplitudes of the processes
$2 \rightarrow 2$ --- one just has to substitute the triple
{\em physical} coupling constants in residues and the
{\em physical} mass parameters in propagators. Therefore the bootstrap
conditions of the second kind obtained at tree level restrict the
allowed values of physical, in principle
{\em measurable} quantities. In this sense
{\em the second kind bootstrap conditions are invariant with respect
to renormalization}, which justifies the legitimacy of the data
analysis presented in
\cite{AVVV1,MENU,Talks}. Note also, that the bootstrap
equations obtained at higher loop orders may differ from the tree
level ones, thus providing
{\em additional} constraints, again, for physical quantities. In fact,
the way we obtained bootstrap in
Sec.~\ref{sec-RP1} ensures that
{\em bootstrap conditions of any loop order present the constraints
imposed on the input parameters by the type of perturbation scheme}.

As we demonstrated in
Sec.~\ref{sec-RP1}, using the first kind bootstrap equations%
\footnote{
They are not needed if the bounding polynomial degrees for the
considered amplitude are negative, like, e.g., in the processes
involving unstable particles.
}
one can obtain well defined expressions for, say, tree level amplitude
in three intersecting hyper-layers. These expressions obey all the
restrictions imposed by fundamental postulates of quantum field
theory. In principle, they can be analytically continued to every 
point of the space of kinematical variables without introducing any 
new parameters. However, there is no guarantee that this analytic
continuation will not generate new singularities in addition to those
already contained in relevant Feynman graphs. Actually, to provide
such a guarantee, is to satisfy the relevant (second kind) bootstrap
equations. In principle, it may turn out that all the bootstrap
equations for processes with
$n \geq 4$ particles at all loop orders should be taken
into account. In other words, the system of RP's
(\ref{eq:RP-tadpole})--(\ref{eq:RP-triple}) is
{\em over-determined} and the question whether the (numerical)
solution exists remains open. We do not discuss this --- extremely
complicated --- problem. Instead, we guess that the solution of all
these bootstrap equations does exist, so that one can consider every
separate equation as a relation between physical observables. In other
words, our results are only concerned with a part of
{\em necessary} bootstrap conditions.

Perhaps, one more detail is noteworthy in connection with the
Eqs.~(\ref{eq:7.05a})--(\ref{eq:7.05b}): to extract the values of
$S$-matrix elements from the full sum
(\ref{7.05}) of
$n$-leg
$l$-loop graphs, one needs to adjust the wave function normalization
constants
$Z$. As we already
mentioned, these RP's are of non-minimal (off-shell) type which we do
not discuss here. Nevertheless, we have to be sure that one does not
need to consider graphs with non-minimal vertices to compute these
constants, or, the same, that our resultant parametrization is
consistent with these non-minimal RP's. In the next Section it is
argued that we need wave function renormalization for stable particles
only, and in
Appendix~\ref{sec-WFren} we show that 
$Z$'s for stable particles are not affected by non-minimal graphs.

\section{Localizability and the fields of resonances}
\label{sec-localizability}
\mbox{}

The main problem we would like to solve (or, at least, to understand
better) is that of constructing the
{\em field-theoretic perturbation scheme} suitable for the case of
{\em strong coupling}, which is closely connected to the
renormalization of canonically non-renormalizable theories. The
results of our previous papers allow us to expect that the effective
field theory concept will be fruitful for finding a solution.

In this Section we discuss the philosophy underlying our approach ---
the concept of
{\em localizable effective scattering theory}. Below we explain this
term and qualitatively describe the relevant
{\em extended perturbation scheme} which, in fact, was considered
in previous Sections. We do not claim to be rigorous here and just try
to give an idea how our technique introduced in
\cite{AVVV2}--\cite{POMI} can be matched with the general ideas of
effective theory formalism.

It is pertinent to mention that the now most popular approach based on
the classical Lagrangian and the canonical quantization procedure
looks impracticable for Lagrangians containing arbitrary high powers
(and orders) of time derivatives of fields: the weight induced in the
functional integral leads to non-hermitian terms in effective
Lagrangian and can be calculated only in relatively simple cases ---
see e.g.
\cite{SalamStrathdee}, or
\cite{WeinMONO}, Chap.~9.3, where relevant calculations are done
for non-linear
$\sigma$-model. That is why we rely upon the alternative ---
intrinsically quantum --- approach proposed in
\cite{WeinQuant} (see also
\cite{WeinMONO}). In that approach the structure of Fock space of
asymptotic states is postulated, and the free field operators are
constructed in accordance with symmetry properties of these states.
This fixes the free Hamiltonian structure. The interaction
Hamiltonian is also postulated as
{\em ab initio} interaction picture operator built from those free
fields and (if necessary) their derivatives.

Strictly speaking, there is no room for unstable particles in
conventional understanding of this scheme. However, as we argue below,
in the framework of effective theory it is convenient to introduce
``fictitious'' resonance fields in order to avoid problems with
divergencies of perturbation series. In a sense, resonances resemble
the ghosts widely known in modern gauge theories. Their fields do not
create asymptotic states and only manifest themselves inside the
$S$-matrix graphs for scattering of stable particles.

For simplicity, in this Section we consider an effective theory which
contains the only field
$\pi$ --- that of massive pseudoscalar particle which we shall refer
to as ``pion''%
\footnote{
Since we do not imply presence of any other symmetry but Lorentz
invariance, this ``pion'' have no isospin.
}.
So, the space of asymptotic states is created by the ``pion'' creation
operator and the free Hamiltonian is just the free ``pion'' energy.

In accordance with the ideology of effective theories, the interaction
Hamiltonian
${\cal H}_I^{\rm initial}$ (in the interaction picture) is constructed
from free ``pion'' field and its derivatives; it contains all the
terms consistent with Lorentz symmetry requirements. No limitation on
the degree and order of time derivatives is implied. The perturbation
scheme for calculating
$S$-matrix elements is based on famous Dyson formula:
\[
S = T_{{}_D} \exp \left( -i \int d^4 x\;  {\cal H}_I^{\rm initial}(x)
\right)\ ,
\]
where
$T_{{}_D}$ stands for so-called Dyson's
$T$-product. By construction, such a theory is renormalizable just
because all kinds of counter\-terms needed to absorb the divergencies
of individual loop graphs are present in the Hamiltonian.

To avoid confusions, we should probably recall one important
circumstance. To ensure Lorentz covariance of the above expression
for the interaction containing field derivatives, one has to include
certain non-covariant ``compensating'' terms in the interaction
Hamiltonian. They cancel the influence of non-covariant propagator
terms appearing due to non-covariance of Dyson's
$T$-product. Fortunately, as one infers e.g. from the discussion in
\cite{WeinMONO} (Chaps.~6.2 and 7.5), the non-covariant terms can be
always thought dropped from both propagator and from interaction
Hamiltonian. That is, we can construct the most general
Lorentz-invariant amplitudes using the most general Lorentz-invariant
interaction Hamiltonian in the Dyson formula written via the
manifestly covariant Wick's
$T$-product (see, e.g.,
\cite{Vasiliev}):
\begin{equation}
S = T_{{}_W} \exp \left(-i \int d^4 x\;  {\cal H}_I^{\rm initial}(x)
\right)\ ,
\label{Texp}
\end{equation}
so that there is no need to take account of any non-covariant terms.

As mentioned in
Sec.~\ref{sec-basic}, the main problem reveals itself when one uses
(\ref{Texp}) for computing the
$S$-matrix elements: the expressions turn out to be purely formal.
Namely, the interaction Hamiltonian contains an infinite number of
items with unlimited number of field derivatives (of arbitrary high
degree and order), so that already at tree level the amplitudes are
represented by infinite functional series. As long as we have no
guiding  principle to  fix the order of summation, we can say nothing
about the sums of such series. Likewise, the higher loop orders are
ill-defined. Therefore it looks natural to single out the special
class of
{\em localizable effective theories} consistent with certain
summability conditions.

Our requirement of localizability can be understood as a system of
restrictions for the Hamiltonian coupling constants. However, as will
become clear from the discussion below, there is no need in explicit
form of those restrictions. Instead, the results of
\cite{AVVV2} allow one to put them in a form of bootstrap equations
for resultant parameters.

To have an idea what localizability is, let us consider the tree level
amplitude of the elastic ``pion-pion'' scattering. At tree level the
relevant part
${\cal H}_{\pi\pi}^{\rm tree}$ of the effective interaction
Hamiltonian
${\cal H}_I^{\rm initial}$ looks as follows:
\[
{\cal H}_{\pi\pi}^{\rm tree} 
= \sum_{i,j=0}^{\infty} G_{ij} \, \pi
(D^{\mu_i}\pi)(D^{\nu_j}\pi)(D_{\mu_i}D_{\nu_j}\pi)\ ,
\]
\begin{equation}
D_{\mu_i} \equiv \partial_{\mu_1} \ldots \partial_{\mu_i}\ ,
\label{Hpipi}
\end{equation}
where
$G_{ij}$ stand for the Hamiltonian coupling constants. It follows from
(\ref{Hpipi}) that the tree level amplitude can be formally presented
as a double series in each pair of independent Mandelstam variables:
\[
A(s,t,u)
= \sum_{i, j} \alpha_{ij} {(s-s_u)}^i {(t-t_u)}^j
\]
\[
= \sum_{i, j} \beta_{ij}  {(t-t_s)}^i {(u-u_s)}^j
\]
\begin{equation}
= \sum_{i, j} \gamma_{ij} {(u-u_t)}^i {(s-s_t)}^j\ ,
\label{treeamp}
\end{equation}
where
${\alpha}_{ij}$, ${\beta}_{ij}$, ${\gamma}_{ij}$,
$s_u$, $t_u$, $t_s$, $u_s$, $u_t$, $s_t$ are some constants which
depend on the pion mass and couplings
$G_{ij}$.

We make an
{\em assumption} (or, better, require) that there is a domain
${\cal D}$ in the space of kinematical variables
$s,t,u$, where at least one of the formal expressions
(\ref{treeamp}) is well-defined (convergent).

For example, suppose that at fixed
$t$ the amplitude in the domain
${\cal D}$ is represented by the converging power series in
$s$ (take
$s_u = 0$ for brevity):
\begin{equation}
A(s,t) = \sum_{i=0}^{\infty} a_i(t) s^i
\label{formalpipi}
\end{equation}
with the coefficients
$a_i$ depending%
\footnote{
The continuity in
$t$ is implied at least in a small interval. Note also that
${\cal D}$ does
{\em not} necessary belong to the physical area of the reaction.
}
on
$t$. However, outside
${\cal D}$ the tree level approximation loses its meaning, leave
alone the higher loop order terms. Similar assumptions are made
about tree level amplitudes of all other processes (elastic and
inelastic).

We see, that in the domain
${\cal D}$ the tree level amplitude is now defined by uniformly
converging power series. As known, the convergency of a power series
is limited by the singularities located on the border of the circle
of convergency%
\footnote{
The radius of this circle should be finite: otherwise the series
would result in polynomially unbounded function not allowed by the
general axiomatic requirements --- see, e.g.,
\cite{axioms}.
}.
The initial power series can not provide a satisfactory description of
the function in the vicinity of singular points. However, if we have
some additional information about these singularities, it may be
possible to single out the singular terms and present our series in a
form that permits the direct analytic continuation to a wider domain.
On the other hand, this new form must also be consistent with
conventional field-theoretic interpretation, otherwise the connection
with basic principles may be lost.

The principles like crossing symmetry, unitarity, etc.\ are saved, if
one finds a kind of ``auxiliary interaction Hamiltonian''
${\cal Q}(x)$ which, when inserted in
(\ref{Texp}), produces the tree level series with the following
properties. First, for every process in initial theory, in every area
${\cal D}$, where tree level series given by (true!) Hamiltonian
${\cal H}_I^{\rm initial}$ converges, the relevant series produced by
${\cal Q}$ shall also converge and result in the same function
(amplitude) --- this is just to be consistent with the initial theory.
Second, these new series shall converge in wider domains thus
providing the analytic continuation of the initial amplitudes. Once
this ``auxiliary Hamiltonian'' results in the tree level series that
converge almost everywhere in the complex space of relevant variables,
the construction of the higher loop order amplitudes is reduced to the
singular integration (in fact, one will need uniform convergence on
compacts, excluding arbitrarily small vicinities of singularities).

Now, if convergent series (of analytical functions, which converge to
an analytic sum) have an ``isolated'' point of divergence (a singular
point, surrounded by convergence area), then at least some items of
the series must also be singular in that point --- that is what
complex analysis says us. Roughly speaking, ``inside'' a domain of
convergence functional series may diverge only in singular points of
its items. So, the ``border'' singularities of the initial tree level
amplitudes should be reproduced by singular terms in the tree level
series generated by
${\cal Q}$. In fact, this is the reason for the summability
requirement formulated in
Sec.~\ref{sec-basic}. Next, the general features of the
field-theoretic formalism reduce severely the types of singularities
in tree level series: only simple poles are allowed. Thereby we also
have to assume (this is one of implicit restrictions that single out
the class of localizable theories) the initial Hamiltonian to be
constructed in a way that
{\em all the singularities of tree-level amplitudes in relevant
variable are just simple poles which can be interpreted as a result of
particle exchange in appropriate channel}. In other words, every
singularity (pole) shall come from a propagator of a (auxiliary)
particle with spin
$J$ and mass
$M$ appearing in the relevant exchange graph of the tree level series
generated by the ``auxiliary Hamiltonian''
${\cal Q}$. Accepting the latter requirement together with summability
principle for loop amplitudes, we can develop the
{\em extended perturbation scheme} based on
${\cal Q}$, rather than
${\cal H}_I^{\rm initial}$.

The set of assumptions (the
{\em localizability hypothesis}) just declared is nothing but an
attempt to develop perturbation scheme in a spirit of famous
quasi-particle method
\cite{Quasi}. In this approach the divergencies of the initial Born
series are cured by introducing a new state (quasi-particle) in such a
way that the corresponding singularity (pole) appears as a single item
in the modified (reconstructed) perturbation series. This method is
widely used in non-relativistic many body problem for potentials that
are too strong to permit the use of perturbation theory over a certain
energy scale. In this approach the fictitious elementary particles are
introduced into the Hamiltonian in correspondence with the spectrum of
bound states of the potential. The potential is also modified, so that
in the domain where the initial Born series converges, the extended
one gives the same transition probabilities. The modified potential is
weaker, since it does not produce those bound states which appear now
as elementary particles. As a result, the modified potential may be
sufficiently weak to allow perturbation methods in a broader energy
domain.

Shortly, the general idea may be formulated as follows: the tree level
amplitude produced by the
{\em extended perturbation scheme} (with auxiliary Hamiltonian
${\cal Q}$) must realize an analytic continuation of the tree level
amplitude obtained in the initial perturbation scheme to a wider
domain. To ensure this requirement all the singularities of the
initial tree level series must be reproduced as separate items of the
new series following from the extended perturbation scheme. Thus the
extended perturbation scheme presents just an auxiliary construction
only needed to properly define the tree level series of the initial
effective theory in a wider domain. Once the tree level amplitude is
defined in the whole space of variables, the calculation of higher
orders of Dyson series becomes just a matter of machinery.

Of course, the feasibility of such an extension strongly depends on
the structure of the initial Hamiltonian. Simply speaking, localizable
Hamiltonian are those for which the described above procedure is
possible. We cannot formulate any
{\em sufficient} condition for existence of such a Hamiltonian, but
we do can investigate the consequences.

It is clear that in the field theory such a modified perturbation
scheme should take account of the interaction of unstable%
\footnote{
Recall that, by construction, all the stable particle states
are already taken into account. This means that the corresponding
poles are present in tree level amplitudes.
}
particles (resonances). It is in this connection that the problem of
field-theoretic description of unstable particles in a finite order
of perturbation theory becomes crucial in our approach. Let us look
closer at how they arise.

The only known way to ensure the consistency of the
$S$-matrix defined by
(\ref{Texp}) with Lorentz covariance requirements and cluster
decomposition principle is to construct the interaction Hamiltonian
density out of free causal quantum fields (see e.g.
\cite{WeinMONO}). The free causal field transforms according to
certain representation of inhomogeneous Lorentz group and coefficients
of its Fourier transform (creation and annihilation operators) satisfy
standard commutation relations which, in turn, uniquely define
corresponding propagator (or, rather, its pole part). Suppose there is
such a field
$\psi_R(x)$, which does not correspond to any asymptotic state of the
initial effective theory: the action of its creation operator on the
Fock vacuum does not form an eigenstate of the free Hamiltonian%
\footnote{
Moreover, the resulting vector does not belong to the initial Fock
space. This will not bother us because we never need to deal with such
vectors: we only need the propagators of auxiliary fields.
}
${\cal H}_0$. Let this field transform according to an irreducible
representation of the inhomogeneous Lorentz group with mass
$M_R$, spin
$J$ and internal quantum numbers of certain
``$\pi \pi$'' channel. Let us formally add to the interaction
Hamiltonian
${\cal H}_I^{\rm initial}$ all possible terms constructed from
$\psi_R$ (and its derivatives) and from the pion fields describing
the true asymptotic states of the initial theory, while all the
coupling constants at these terms are treated as free parameters. The
obtained operator%
\footnote{
In our previous papers we called it as ``extended Hamiltonian'', which
is not quite correct. The point is that this operator cannot be
obtained as a result of transition to the interaction picture of the
Heisenberg picture Hamiltonian describing pions. Nevertheless, in what
follows we sometimes use the term ``extended Hamiltonian'' just for
the sake of brevity.
}
looks as follows:
\[
{\cal Q}(x)=
{\cal H}_I^{\rm initial}(x)+{\cal H}_I^{\rm extra}(x).
\]
Note also, that
{\em the free Hamiltonian structure remains unchanged}. Consider now
the extended perturbation scheme, in which the operator
${\cal Q}(x)$ replace the Hamiltonian in the matrix elements of
(\ref{Texp}) between the asymptotic states of the initial effective
theory (containing only ``pions''). Hence, in addition to diagrams of
the initial theory there will appear diagrams with internal exchanges
by (fictitious) particles
$\psi_{R}$. It is crucial for us that the new tree level amplitude of
$\pi \pi$ scattering will have an explicit pole at
$s=M_R^2$ (as well as, of course, the poles in
$t$ and $u$). Now, to require the localizability is to require that
the new tree level amplitude with explicit pole part is equal to the
initial tree level amplitude
(\ref{formalpipi}) in the domain
${\cal D}$:
\[
A(s,t) =
\sum_{i=0}^{\infty} b_i(t) s^i + \frac{r(t)}{s-M_{R_s}^2}\
=
\sum_{i=0}^{\infty} a_i(t) s^i 
\]
\begin{equation}
{\rm for\;} (s,t)\in{\cal D},
\label{reorgpipi}
\end{equation}
but the new series is valid in a wider domain%
\footnote{
For definiteness we assumed that it is the
$s=M_R^2$ pole lying at the border of convergency circle of initial
series
(\ref{formalpipi}).
}.
This equation can be interpreted as a system of
{\em matching conditions} for the residue
$r(t)$ and the coefficients
$b_i(t)$ of new series, which, in turn, can be expressed through the
coupling constants of new ``Hamiltonian''
${\cal Q}$ and the quantum numbers of filed
$\psi_R$.

In principle, one can imagine certain step-by-step extension of
perturbation scheme with the resonance fields introduced in accordance
with singularities that limit the convergency of the initial tree
level series for amplitudes. These steps should be repeated until the
tree level amplitude is defined everywhere in the space of kinematical
variables. However, there is another way which looks much more
promising. One can consider the most general extended perturbation
scheme containing an arbitrary number of resonances of arbitrary high
spin and mass and then analyze the general structure of such a scheme%
\footnote{
A similar ideology is used, for example, in
\cite{Bogoliubov} in perturbation theory for non-linear oscillations.
}.
This is precisely the way which we follow in our studies.

Employing few additional principles (summability and uniformity, see
Sec.~\ref{sec-basic}), we have shown that the structure of such
perturbation scheme is far from being arbitrary. The minimal
parametrization described in
\cite{AVVV2} proved to be especially convenient for the classification
of coupling constants appearing in calculations. Besides, as shown in
Secs.~\ref{sec-RP1}--\ref{sec-RP3} (for preliminary discussion see
\cite{POMI}), the necessary conditions of self-consistency of the tree
level approximation impose strong limitations on the possible set of
the parameters of extended perturbation scheme. Altogether, this gives
a hope that the concept of localizable effective theory may prove to
be fruitful in studies of the resonance physics.

Let us make a summary, just repeating what have been said about the
localizability:
{\em the effective theory is called  localizable if, first, its formal
series for all the tree level amplitudes converge at certain small
domains in the spaces of kinematical variables and, second, in those
domains they can be reproduced by well-defined in a wider domains tree
level series of the extended perturbation scheme containing auxiliary
resonance fields with spins
$J_i$ and ``masses''%
\footnote{
Resonance masses are discussed in Section
\ref{sec-resonances} below.
}
$M_i$ which do not belong to the spectrum of one-particle asymptotic
states}. The auxiliary fields may be interpreted as those describing
particles which are unstable with respect to decays into the true
asymptotic states of initial theory.

One of the immediate consequences of this is that in the extended
perturbation scheme the only
$S$-matrix elements one needs to compute are those between the true
asymptotic states%
\footnote{
In this respect the philosophy of the approach based on the extended
perturbation scheme is quite consistent with that developed in
\cite{Veltman}.
}.
The normalization condition remains the same as that in the initial
effective theory and the unit operator
$\hat{\bf 1}$ in the Fock space takes a form of infinite sum over the
states only containing stable particles:
\[
\hat{\bf 1}
= \sum_{n=0}^{\infty}
|n\;
({\rm \scriptstyle stable!})\rangle
\langle n\; ({\rm \scriptstyle stable!}) |.
\]
Therefore the only RP's needed to provide the correct normalization of
the wave functions are those for the self-energy graphs of stable
particles (``pions'' in the case in question). As mentioned in
\cite{AVVV2}, the computation of relevant counterterms does not
require operating with non-minimal vertices. The proof of the latter
statement is given in
Appendix~\ref{sec-WFren}.

The last, but very important remark to be made here is that explaining
our usage of term ``strong coupling'' or ``strong interaction''. Let
us call the interaction described by certain localizable effective
theory as
{\em strong}, if the number of steps needed for complete localization
is actually
{\em infinite}. To put it another way, in the case of ``strong''
coupling one needs an infinite number of resonances to construct the
extended perturbation scheme: it is not possible to point out the
upper limit for
$M_i$. Surely, we imply that the number of resonances falling into
arbitrary finite energy interval is finite, so that  the tree level
amplitudes of strong processes belong to the class of meromorphic
functions. In contrast, the theories which require just a finite
number of localization steps produce the tree level amplitudes
expressed by the rational functions having a finite number of poles.
This latter interaction type we may call as weak. The immediate
example of such a theory is the celebrated electro-weak model
considered at energies below the mass of Higgs or Z-boson  (depending
on which one is taken to be lighter).

The important physical difference between strong and weak interactions
is that in the strong case the degree of the bounding polynomial for
an amplitude depends on the hyper-layer under consideration. It is
this kind of amplitudes that our technique is developed to handle. On
the other hand, it is easy to understand that in a theory with finite
number of resonances (weak case) the amplitude asymptotics is given by
the highest degree of kinematical variable present, and, hence, does
not depend on the layer. So,
{\em perhaps, the most interesting information which can be extracted
from hadron experiments is that allowing to fix the high energy
asymptotics of the amplitudes of exclusive processes at various fixed
values of the other kinematical variables}.

\section{On the parametrization of resonance}
\label{sec-resonances}
\mbox{}

The way resultant parameters were introduced in
\cite{AVVV2} and the way we impose renormalization prescriptions in
Sec.~\ref{sec-RP3} assume the use of
{\em renormalized perturbation theory} with
{\em on-mass-shell renormalization conditions}. So, when we speak
about the stable particles (like e.g. pion or nucleon), the mass is
the quantum number of corresponding asymptotic state, and it is the
same quantity that appear in the Feynman propagator. In contrast, the
terms ``mass'' and ``width'' often used to describe a resonance are
not so well defined. Here we shall trace their relation to the
parameters used in our approach.

The most important fact proven in
\cite{AVVV2} is that every
$N$-loop
$S$-matrix element can be represented as a sum of graphs constructed
from resultant vertices. The proof implies that each amplitude is
calculated precisely at a given loop order and no kind of partial
re-summation (like, e.g., the Dyson re-summation of the propagator) is
allowed. Hence, at any finite loop order, every pole of the amplitude
is a simple pole at real values of the momentum squared. This means
that the customary Breit-Wigner description of resonances (mass and
width as real and imaginary parts of the pole position, respectively)
looses
its meaning in terms of the finite level resultant parameters%
\footnote{
Actually, this is just a consequence of the fact that the notion of
resonance is ill-defined in the framework of perturbative quantum
field theory.
}.
This creates a problem when one needs to compare the (finite loop
order) theoretical expression with experimental data.

One of the ways to circumvent this difficulty is to exclude the small
vicinities of resonances from the data analysis. Technically, this is
just a problem of appropriate organization of the fitting procedure;
it can be easily solved. In contrast, from the purely theoretical
viewpoint there is a problem of
{\em suitable definition for the parameters describing a resonance}.
We shall stress that the latter
{\em may only make sense in the framework of a particular perturbation
scheme. If we could construct the complete non-perturbative
expressions for
$S$-matrix elements we would never need to use this term}.

As explained in
Sec.~\ref{sec-RP3}, throughout the article the term
``physical mass'' refers to the parameter
$M_i$ in the denominator of the (stable or unstable) particle
propagator. The conventional term ``width'' is closely related to our
definition
(\ref{eq:phys-coupling}) of ``physical coupling constants''. That is,
when fitting data with the finite loop order amplitude one obtains the
values of relevant physical couplings which, in turn, can be used to
compute (formally!) the decay amplitude and, hence, the resonance
widths. Therefore, in principle, it is possible to avoid using the
term ``width'' by operating only with the values of physical coupling
constants and mass parameters.

However, when comparing our bootstrap equations with experiment we are
forced to use the numbers quoted, say, in
\cite{PDG}. Those numbers are obtained by data fitting with amplitudes
constructed in terms of
{\em complex} poles and smooth background. For example, in
\cite{PDG} the resonance characteristics are given in terms of mass
and width, which are the parameters of the T-matrix poles at an
unphysical sheet of the complex energy plane. In many other sources
the data on resonances are quoted in terms of Breit-Wigner or K-matrix
parameters. One should realize, that these phenomenological constants
never appear in the finite loop order expressions for field-theoretic
perturbative amplitudes produced by the formula
(\ref{Texp}). Moreover, there is no guarantee at all that the partial
sums of Dyson series (or any other partial summation) can provide a
reasonable sequence of approximants for the amplitude at
{\em complex} values of
kinematical variables: this is a purely mathematical problem still
awaiting a solution. All this means that the most reliable way to
verify predictions of field-theoretic models is to fit data with the
parameters that appear in field-theoretic formulae. The use of any
other kind of parametrization (say, Breit-Wigner plus background) may
lead to uncontrollable errors. In particular, there is no direct
correspondence between the Breit-Wigner resonance parameters (mass and
width as real and imaginary parts of the pole position) and discussed
above field-theoretic ones (mass as the Feynman propagator pole
position and width as the value of corresponding ``decay amplitude''
calculated in terms of physical coupling constants at a given loop
order).

Nevertheless, the use of numerical data from
\cite{PDG} for approximate checking of selected bootstrap equations
(see
\cite{MENU,Talks}) looks quite justified, at least for rough
estimates. Indeed, for well-separated narrow resonance the different
methods of parametrization (including those based on field theory
models operating directly with coupling constants) result in
approximately the same values of masses and couplings. However, this
is not true for broad resonances like famous light scalars%
\footnote{
An excellent discussion of this point presented in the series of
papers
\cite{Schechter}.
}.
In this latter case the values of mass and coupling constant extracted
from the same data set with the help of different theoretical
amplitudes may differ considerably. To avoid inconsistencies, in
\cite{MENU,Talks} we only rely upon the data
\cite{PDG} as long as considered bootstrap equations are well
saturated by a set of relatively narrow resonances, while the
remaining equations are used to make certain estimates and
predictions.

\section{Effective theory versus analytic
         $S$-matrix}
\label{sec-analyt}
\mbox{}

A wide use of complex analysis and some terminological analogies may
turn the reader to think that the physical ideas underlying our
approach are similar to those widely known as analytic
$S$-matrix philosophy (see, e.g.
\cite{Chew}). It is not so, and here we discuss the main differences.

First, the analytic
$S$-matrix approach (ASM) is based on the idea of
{\em nuclear democracy}: all the particles (together with all their
composites like nuclei) are considered on the same footing. For this
reason the methods based on the Hamiltonian (Lagrangian) were rejected
by ASM ideologists. In fact, neither Fock space nor Dyson's formula
for the
$S$-matrix play a role there. In contrast, our scheme uses the
field-theoretic --- Dyson's --- construction of
$S$-matrix as the operator acting on the Fock space of true asymptotic
states. These states are created by field operators that describe just
``aristocratic'' --- stable species of particles (pions and nucleons
in
$SU_2$ sector). It is implied that the quantum numbers (mass, spin,
etc.) of these particles can be explained by some underlying
fundamental theory (say, QCD). For us those numbers are just external
parameters which (perhaps, along with few another ones) should be
taken from experiment. These and only these particles (or, better,
the corresponding operators) appear in the free Hamiltonian%
\footnote{
This statement is not quite correct. Along with pion and
nucleon terms one should add to free Hamiltonian also the items that
correspond to their
{\em stable} composites (nuclei). For now we just close our eyes to
the existence of such objects.
}.
So, from the very beginning we do not consider resonances on the same
ground as stable particles. In our approach the resonance fields only
appear in the structure of the
{\em extended interaction Hamiltonian}
${\cal Q}$: they are nothing but a convenient tool for presenting the
analytic continuation of tree level amplitudes in the form consistent
with Dyson's formula
(\ref{Texp}). In other words,
{\em in our approach the free Hamiltonian is assumed to have the same
spectrum as that of the full one}. It is this feature which gives us a
hope to develop the efficient field-theoretic perturbation scheme.

The main reason why we prefer to deal with
$S$-matrix elements and not with Green functions is that in the former
case we can point out the full set of numerical parameters that appear
in the process of perturbative calculations (resultant parameters).
Besides, as stressed in
\cite{WeinAsySafe}, the essential parameters (which can be built of
resultant ones) play a special role in canonically non-renormalizable
theories. It is interesting to note that our system of RP's
(\ref{eq:RP-tadpole})--(\ref{eq:RP-triple}) looks precisely like that
conventionally used in ordinary renormalizable theories (like, say,
${\phi}^3$), though, of course, in our case the number of field
species is infinite. This result checks well with Weinberg's
conjecture made in Ref.~\cite{WeinAsySafe}.

In fact, all the other distinctive features of the approach based on
the effective theory concept are just consequences of Dyson's formal
construction of the
$S$-matrix. The localizability, as well as uniformity and summability
requirements
(Secs.~\ref{sec-localizability} and
\ref{sec-basic}) are only needed to handle individual terms of Dyson
series in mathematically correct way. These requirements allow one,
first, to single out the set of essential parameters and, second, to
put this set in order of increasing loop level. At last, the natural
self-consistency conditions (that mirroring the requirement of
crossing symmetry at a given loop order) make it possible to
understand the origin of regularity known as bootstrap restrictions%
\footnote{
As far as we know, the explicit form of those restrictions or, at
least, of their part has been never obtained in the framework of ASM
approach.
}.

One of the most important technical differences between the philosophy
of ASM and the effective field theory approach is that in the former
case the coupling constants at a given vertex are considered
independent of the loop order under consideration. As we have shown,
this is not so in the framework of effective theory. Instead, it is
the dependence of coupling constants on the loop order that --- as
shown in
Secs.~\ref{sec-RP2} and
\ref{sec-RP3} --- makes it possible to write down the explicit
form of the lowest (tree) level bootstrap conditions for RP's. The
fact that these conditions restrict
{\em physical} parameters is a direct consequence of the resultant
parametrization, which implies the use of renormalized perturbation
theory. So, while we never know how many loops in the Dyson series
should be considered to get reasonable coincidence with experimental
amplitude, the bootstrap equations, when written in the form
(\ref{4.9}), are
{\em exact} equalities at any loop order, though, of course, in
numerical tests one is limited by the set of established resonances
and experimental precision.

\section*{Conclusion}
\label{sec-conclusion}
\mbox{}

At first glance, the concept of effective theory might seem too
general to be of practical use in computing the amplitudes of strong
processes. However, this is not quite true. The famous Chiral
Perturbation Theory (see
\cite{WeinbEFT}, \cite{ChPT})
provides an example that disproves this opinion. Unfortunately, the
problem of infiniteness of the required set of renormalization
prescriptions, the inseparable feature of non-renormalizable theories,
remains a stumbling block for this approach. The above-discussed
concept of localizable effective scattering theory provides a way to
solution. The localization procedure along with subsequent reduction
of all the lines of relevant graphs allow one to single out the
well-ordered (though still infinite) subset of minimal RP's needed for
complete renormalization of
$S$-matrix
elements. In physically interesting cases this subset only contains the
RP's for 1-, 2-, 3- and, possibly, 4-leg minimal vertices. Moreover,
as shown above, the prescriptions collected in this set cannot be
chosen arbitrarily because they must fulfil the infinite number of
bootstrap conditions. So, the question on the full number of
{\em independent}
prescriptions takes on great importance. Unfortunately, it still
remains unanswered. At the moment we only can say that the number of
independent RP's needed for complete renormalization of a given
effective scattering theory equals that of solutions of the full
system of second kind bootstrap conditions. In the next paper we will
show that the set of independent non-minimal prescriptions is governed
by similar bootstrap equations.

In this connection it is pertinent to stress that Gross's note
\cite{Gross}
on the essential infiniteness of independent solutions of the
bootstrap constraints only relates to the case of finite-component
theories. According to the definition in
Sec.~\ref{sec-localizability},
the theories of such a kind correspond to weak interaction. They
cannot describe the situation when the asymptotic behavior (more
precisely, the bounding polynomial degree) of an amplitude depends on
the momentum transfer. One of the most important properties of strong
interaction is that the degrees of bounding polynomials (which
characterize the high energy behavior) are strongly correlated with
the values of remaining kinematical variables. Perhaps, it is this
correlation which could provide a key idea on the general structure of
the full system of bootstrap conditions and, hence, to elucidate the
question on its solvability and the full number of solutions. This is
a problem for the future investigations. As to the results obtained
thus far, it seems us most important to carry out the systematic
approximate comparison of known experimental data with the second kind
bootstrap conditions for the parameters of resonances appearing in
concrete elastic scattering processes. This will be done in two
separate publications on pion-nucleon and kaon-nucleon reactions.

\begin{acknowledgments}
\mbox{}

We are grateful to V.~Cheianov, H.~Nielsen, P.~Osland, S.~Paston,
J.~Schechter, A.~Vasiliev and especially to M.~Vyazovski for
stimulating discussions. The work was supported in part by INTAS
(project 587, 2000) and by Ministry of Education of Russia (Programme
``Universities of Russia''). The work by A.~Vereshagin was supported
by L.~Meltzers H\o yskolefond (Studentprosjektstipend, 2004).

\end{acknowledgments}

\appendix

\section{The Cauchy Forms }
\label{sec-AppCauchy}
\mbox{}

The existence of the Cauchy form for a
{\em meromorphic} (no cuts, only poles) polynomially bounded function
is a special case of the famous Mittag-Leffler theorem valid for any
meromorphic function of one complex variable
\cite{Complex}. We used this representation in a way adjusted for two
variables when working with tree level amplitudes in
\cite{AVVV1}. To have a feeling how this technique works with concrete
mathematical examples one is addressed to
\cite[Sec.~4]{AVVV1}, where it is shown that the commonly accepted
interpretation of so-called ``duality hypothesis''%
\footnote{
The absence of contributions from cross-channel poles and point-like
vertices in the direct channel amplitude, see e.g.
\cite{dual}.
}
is not correct even in the case of Veneziano model
\cite{Veneziano}. The bootstrap equations for that model obtained with
the help of Cauchy forms are further explored in
\cite{POMI}. The Cauchy form (or, the same, Cauchy series) can be
easily derived from the
Eq.~(\ref{3.2}) of
Section~\ref{sec-Cauchy}, that is why we found it natural to discuss
them here.

Consider the situation when all the singularities
$s_k({\bf x})$ of the
$N$-bounded function
$f (z, {\bf x})$ investigated in that Section are just
{\em simple poles}. Hence, the cuts are not needed anymore and all
the contours
$C_k$ in
Fig.~\ref{fig:sys-of-con} are transformed into small circles around
the poles. Let us denote the residue at the pole
$s_k$ as
$r_k ({\bf x})$. Using the residue theorem, the integrals on the right
side of
Eq.~(\ref{3.2})
are easily performed and the expression takes the following form:
\[
f(z, {\bf x})
= \sum \limits_{n=0}^{N}
\frac{1}{n!} f^{(n)}(0, {\bf x}) z^n
\]
\begin{equation}
+ \sum_{k=-\infty }^{+\infty}
\left\{
\frac{r_k({\bf x})}{z-s_k({\bf x})} - h^{N}_k({\bf x},z)
\right\}\ ,
\label{capp3}
\end{equation}
where the
{\em correcting polynomials}
\begin{equation}
h^{N}_k (z, {\bf x})
\equiv -\frac{r_k({\bf x})}{s_k({\bf x})} \,
\sum\limits_{n=0}^N\, {\left[ \frac{z}{s_k({\bf x})} \right]}^n
\label{capp4}
\end{equation}
ensure the convergence of resulting series. We have stressed in
Sec.~\ref{sec-Cauchy} that one should sum the items in this series in
order of increasing
$|s_k|$. As long as this rule is kept, the series converge uniformly
in the layer
$B_x$. Once this rule is violated or correcting polynomials are
dropped, the convergence may be lost%
\footnote{
There are exceptions --- see the next footnote.
}
and the series loses its meaning. This is, of course, governed by the
estimate
(\ref{3.1}).

The relation
(\ref{capp3}) is precisely what we call the Cauchy form (series) for
the function
$f(z, {\bf x})$. In
\cite{AVVV1} we called the first term on the r.h.s. of
(\ref{capp3}) as the
{\em external part}, while the first terms in curly brackets are
conventionally called the
{\em principal parts} of the function.

When
$N$ becomes negative, both the correcting polynomial and external part
disappear, and the Cauchy form is reduced to the simple sum of
principal parts --- the pole terms. More generally, it is easy to show
that for the
$N$-bounded in the layer
$B_x$ function presented by
(\ref{capp3}) certain ``collapsing'' conditions are valid: the
correcting polynomial degrees higher than
$N$ converge separately to the values of corresponding derivatives:
\[
\sum_{k= -\infty}^{+\infty}
\frac{ r_k({\bf x}) }{ \left[ s_k({\bf x}) \right]^{N'+1} }
= \frac{1}{N'!}f^{(N')}(0, {\bf x})\ ,
\]
\begin{equation}
N'= N+1, N+2, \ldots\ .
\label{capp5}
\end{equation}
To put it another way, suppose that one uses some higher degree
$N' > N$ in the
Eq.~(\ref{capp3}). Since estimate
(\ref{3.1}) holds for
$N'$, the Cauchy form certainly converges but can be reduced to one
with true degree
$N$ --- the superfluous degrees of external part and correcting
polynomials just cancel. So, roughly speaking, the appearance of
Cauchy's form is strongly correlated with asymptotics.

In case of physical interest the function
$f$ represents the tree-level amplitude of a process, while
$z$ and
$x$ play a role of kinematical variables. At the first glance at
Eq.~(\ref{capp3}) it is tempting to separate the correcting
polynomials from contribution of the poles, sum up two resulting
series independently and combine the sum of correcting polynomials
with the external part. This could give a polynomial in
$z$ plus a sum of poles. However this is
{\em not correct} until the amplitude
$f$ possesses the decreasing asymptotics
$N < 0$. Instead, as long as
Eq.~(\ref{3.1}) does not hold for negative
$N$, the sum of poles and the sum of correcting polynomials taken
separately
{\em diverge}%
\footnote{
Except the trivial case when poles do not take part in forming the
non-decreasing asymptotics which, thus, turns out to be caused by the
properties of external part. Then the infinite series of poles
converges independently, as well as the series of correcting
polynomials. Such a situation does not present any interest for
constructing the effective theory of strong interactions because, as
shown in
\cite{AVVV1}, it does not occur even in Veneziano model.
},
and the re-summation procedure is illegal. Therefore, in case if
$N \geq 0$, there is a
{\em crucial difference} between approximating some amplitude by a
polynomial plus a finite sum of pole terms (as it is often done when
the amplitude is saturated by resonances), and approximating it by the
finite sum of terms of the corresponding Cauchy form. The first way is
nothing but an attempt to approximate by the first few terms of the
divergent series and balance the situation by adjusting the
polynomial. Therefore, as it should be with the divergent series,
every next pole term taken into account forces one to change the
polynomial significantly. This leads to instability of approximation.
In contrast, this does not happen if one uses first terms of the
relevant Cauchy series with correct asymptotics, because these series
converge by construction.

\section{The case of non-decreasing asymptotics}
\label{sec-Appasymp}
\mbox{}

For completeness, here we consider the situation when the process
$2 \rightarrow 2$ is described by the amplitude (or, more precisely,
scalar formfactor) characterized by non-negative degrees of the
bounding polynomials in each of the three layers:
$B_s$, $B_t$ and
$B_u$ (see Fig.~\ref{fig:triangle}), though, as we have noted in
Sec.~\ref{sec-RP3}, such a process (with the hadrons stable w.r.t.
strong decays) is forbidden by known phenomenology. As mentioned in
Sec.~\ref{sec-RP1}, in addition to the renormalization prescriptions
for 3-leg vertices, we will also need here RP's for 4-leg one. What is
significant, is that in this case to ascertain the number (and the
form) of required RP's we will need to employ the bootstrap conditions
of the second kind.

To be specific, let us take
$N_t = N_u = 0$ (we use the same notations as those in
Sec.~\ref{sec-RP1}). Hence, in the layer
$B_t$ the amplitude
$M$ of the process can be presented as
\begin{equation}
M\Bigr|_{B_t} = f(t) + F(t,{\nu}_t)\ ,
\label{A1.1}
\end{equation}
where
$F(t,{\nu}_t)$ is an infinite sum of contour integrals appearing in
(\ref{4.3}) and
$f(t)$ is unknown function only depending on
$t$. Similarly, for the same amplitude in
$B_u$ we have:
\begin{equation}
M\Bigr|_{B_u} = \phi (u) + \Phi (u,{\nu}_u)
\label{A1.2}
\end{equation}
with another unknown function
$\phi (u)$ (only depending on
$u$) and an analogous sum of contour integrals
$\Phi (u,{\nu}_u)$. Recall that each of the above expressions is well
defined (convergent) only in the corresponding hyper-layer. Both forms
(\ref{A1.1}) and
(\ref{A1.2}) are written for the same amplitude, therefore in
$D_s \equiv B_t \cap B_u$ they must identically coincide with one
another:
\begin{equation}
f(t) + W (t,u) = \phi (u)\ , \quad (t,u)\in D_s\ ,
\label{A1.4}
\end{equation}
where
\[
W(t,u) \equiv F(t,{\nu}_t) - \Phi (u,{\nu}_u)\ .
\]
The self-consistency requirement
(\ref{A1.4}) provides us with a source of an infinite system of
bootstrap conditions.

Let us first derive the conditions of the second kind. From
(\ref{A1.4}) we obtain
\begin{equation}
{\partial}_t {\partial}_u W(t,u) \equiv 0\ , \quad
(t,u)\in D_s\ .
\label{A1.5}
\end{equation}
According to the results of
Sec.~\ref{sec-RP2}, both series
$F(t,{\nu}_t)$ and
$\Phi (u,{\nu}_u)$ should be considered as known functions completely
determined by the resultant parameters fixed on the previous steps of
renormalization procedure. Then in
$D_s$ their difference
$W (t,u)$ is also well-defined known function%
\footnote{
Using the equality
$s+t+u= \sum\limits_{i=1}^4 m_i^2\;$ ($m_i$'s are the masses of
external particles), express
${\nu}_t \equiv s - u$ and
${\nu}_u \equiv t - s$ via
$t$ and
$u$.
}
of
$t$ and
$u$. Hence, the subsystem
(\ref{A1.5}) restricts the allowed values of those fixed resultant
parameters or, the same, restricts the allowed values of the relevant
RP's.

Now, differentiating equation
(\ref{A1.4}) separately w.r.t
$t$ or
$u$ we get:
\[
{\partial}_t f(t) + {\partial}_t W(t,u) = 0\ , \quad
{\partial}_u \phi(u) - {\partial}_u W(t,u) = 0\ ,
\]
which, when solved, gives:
\[
f(t)     = -  W(t,u) + W(t_0,u) + f(t_0)\ , 
\]
\begin{equation}
\phi (u) =    W(t,u) - W(t,u_0) + \phi (u_0)\ , 
\label{A1.7}
\end{equation}
for any
$(t_0,u_0) \in D_s$. Substituting these relations into
(\ref{A1.4}) one obtains:
\begin{equation}
f(t_0) =
\left[ W(t,u) - W(t,u_0) - W(t_0,u) \right] + \phi (u_0)\ .
\label{A1.8}
\end{equation}
On the other hand, again from
(\ref{A1.4}), we know that
\begin{equation}
f(t_0) = - W(t_0,u_0) + \phi (u_0)\ .
\label{A1.9}
\end{equation}
Combining
(\ref{A1.8}) with
(\ref{A1.9}) we obtain the universal form of the subsystem of second
kind bootstrap conditions:
\[
W(t,u) = W(t,u_0) + W(t_0,u) - W(t_0,u_0)\ , 
\]
\begin{equation}
{\rm for \;}
(t,u) \in D_s\ ,\; (t_0,u_0) \in D_s\ ,
\label{A1.10}
\end{equation}
which could also be derived directly from
(\ref{A1.5}). The forms of this type always appear during analysis of
the bootstrap conditions in the intersection of two hyper-layers.

We can now answer the question on how many RP's are needed to fix the
amplitude under consideration. From the relations
(\ref{A1.7}) it follows that both functions
$f(t)$ and
$\phi(u)$ are known if we specify the values
$f(t_0)$ and
$\phi(u_0)$. Further,
(\ref{A1.9}) tells us that it is enough to know one of the latter
constants. Hence, we need to add only one additional RP to the system
(\ref{eq:RP-tadpole}--\ref{eq:RP-triple}). It is natural to fix the
amplitude value at one point arbitrary chosen either in
$B_t$ or in
$B_u$. For example, one may do it in
$B_u$ as follows:
\[
M \Bigr|_{u = 0;\; \nu_u = 0} =  g_{00}\ ,
\]
where
$g_{00}$ may be considered as a parameter given, say, by fit with
experimental data. The latter equality, in principle, can be
understood as an indirect fixing of corresponding 4-leg resultant
vertex in that point of variable space. However it is not easy to
trace the contribution of that vertex to the series
(\ref{A1.2}), that is why an explicit fixing may be tricky and
inconvenient for calculations.

We conclude that in the case of ``constant asymptotics'' in both
intersecting layers
($N_t = N_u = 0$) one needs to add only one new renormalization
prescription for 4-leg amplitude, in addition to those for 3-leg
couplings considered in
Sec.~\ref{sec-RP2}. Note also, that it may happen convenient to fix an
{\em amplitude} value in some chosen point, as we just did, and not
directly the value of relevant resultant 4-leg vertex.

By the same method one can examine an amplitude with arbitrary high
asymptotic in both layers. In particular, for the physically
interesting situations
($N_t = 1$, $N_u = 0$) and
($N_t = 1$, $N_u = 1$) one will need to fix two and three numbers,
respectively. The corresponding RP's can be written, say, as follows:

\noindent
$(N_t = 1,\; N_u = 0):$
\[
M \Bigr|_{u=0;\;\nu_u=0} = g_{00}\ , \quad
\partial_u M \Bigr|_{u=0;\;\nu_u=0} = g_{10}\ ;
\]

\noindent
$(N_t = 1,\; N_u = 1):$
\[
M\Bigr|_{u=0;\;\nu_u=0} = g_{00}\ , \quad
\partial_u M\Bigr|_{u=0;\;\nu_u=0} = g_{10}\ , 
\]
\[
\partial_u \partial_{\nu_u}
M\Bigr|_{u=0;\;\nu_u=0} = g_{11}\ .
\]

\section{Self-energy graphs}
\label{sec-WFren}
\mbox{}

In
\cite{AVVV2} it was said very little about the wave function
(field-strength) renormalization. In this Appendix we explain why
the computation of the wave function renormalization constants does
not require operating on non-minimal graphs%
\footnote{
As usual the absence of massless particles with spin
$J \geq 1$ is implied.
}.
To put it another way, when calculating the S-matrix one can simply
neglect the residual graphs with non-minimal
{\em external} lines that certainly remain after the reduction
described in
\cite{AVVV2}.

\begin{figure}[!ht]
{\bf I.}
\includegraphics{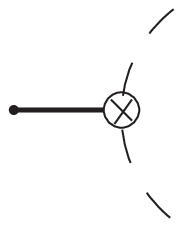}
\hspace{1cm}
{\bf II.}
\includegraphics{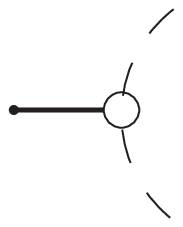}
\caption{
Graph after reduction: some vertices, non-minimal w.r.t.\
corresponding external lines, can remain (case ``II''). The rest of
the graph (dashed circle segment) contains only minimal vertices.
\label{fig:ext-line}
}
\end{figure}

The problem is that this reduction produces graphs of two different
kinds. All vertices of the first kind graphs are minimal w.r.t.\
internal and (if connected to) external lines. Therefore these graphs
depend only on minimal parameters:
Fig.~\ref{fig:ext-line} (I).
In contrast, at least one vertex connected with external line of a
second kind graph is non-minimal:
Fig.~\ref{fig:ext-line} (II). All the internal vertices are minimal
in both cases, and, according to the footnote in
Sec.~\ref{sec-prelim}, all the propagators are always taken minimal.
Surely, every graph of the second kind vanishes when viewed as a part
of an
$S$-matrix element because the latter is always computed on the mass
shell. However, it is not a good reason to drop these graphs out of
consideration: their parameters may still contribute to the wave
function renormalization constants. It may seem then that to take
account of this effect one needs to calculate the second kind graphs
and, hence, to deal with non-minimal parameters. In other words, the
renormalization programme discussed in
Secs.~\ref{sec-RP1}--\ref{sec-RP2} fails. Fortunately, this is not the
case. The reason was pointed out in
\cite{AVVV2}, here we only would like to discuss details.

As argued in
Sec.~\ref{sec-localizability}, in the extended perturbation scheme one
never needs to compute
$S$-matrix elements between the states with unstable particles as the
Fock space is created by free field operators of the stable particles
only. Technically it means that there is no need in the ``wave
function'' renormalization for resonances and only field-strength
renormalization constants for true asymptotic states may be of
interest. Therefore we can safely refer to the conventional derivation
of the Lehmann-Symanzik-Zimmermann (LSZ) reduction formula%
\footnote{
See, e.g.
\cite{WeinMONO}, \cite{Peskin}; it should not be mixed with the
reduction to minimal form (reduction procedure or reduction theorem
\cite{AVVV2}) that we discuss here.
}.

\begin{figure}[!ht]
\mbox{
{\bf m.}
\includegraphics{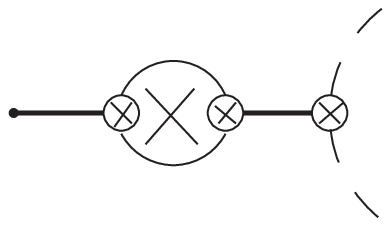}
}
\mbox{
{\bf n.}
\includegraphics{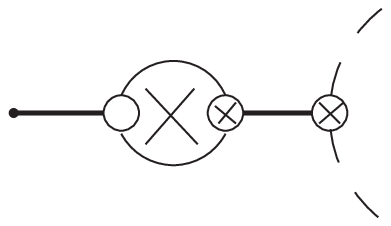}
}
\caption{
Self energy insertion in external leg of a reduced graph. Minimal
vertices are indicated by crosses: only the very left vertex can be
non-minimal (case
``{\bf n}'').
\label{fig:sigma}
}
\end{figure}

The field-strength renormalization is caused by 1PI self-energy
(2-leg) insertions in external lines of graphs that contribute to a
given scattering process. Suppose that the reduction is done and,
hence, all the vertices are minimal w.r.t. internal lines. External
line without insertions can be amputated and put on-shell immediately,
thus the relevant vertex can always be taken minimal. So, let us look
at the external leg with self-energy insertions. According to
\cite{AVVV2}, we can then distinguish two possibilities:
Fig.~\ref{fig:sigma}.
The first one suits us, since relevant 2-leg graph contains only
minimal vertices
({\bf m}-type graphs). The second is uncomfortable since the very left
vertex (that without cross) is non-minimal
({\bf n}-type graphs).

To see that the latter insertion does not affect the
$S$-matrix, we recall that the only part of two-point function
contributing to the LSZ formula is the residue in the one-particle
exchange pole, proportional to the wave function renormalization
constant
$Z$ (see e.g.
\cite{Peskin} Chaps.~7.1--7.2). Hadronic phenomenology says us that
the highest spin we may need for external (stable) particle is
$\frac{1}{2}$. So let us for definiteness work with nucleon
propagator. The analysis below (we do not discuss mixing) can be
easily adjusted for any spin, the scalar case is much simpler. The
isospin structure is irrelevant for current discussion and, therefore,
will be neglected.

For spin-%
$\frac{1}{2}$ particle the (parity conserving) amputated 1PI two-point
function
$\Sigma (p)$ (that contains all the contributions of both
{\bf m} and
{\bf n} types) can always be expressed via matrices
$\hat{\bf 1}$ (the unit matrix) and
$\s{p}$ as
\[
\Sigma (p) = \sigma_1(\pi) \hat{\bf 1} + \sigma_2(\pi) (\s{p}-m)\ ,
\quad \pi \equiv p^2 - m^2\ ,
\]
so that the full propagator takes the form:
\[
\raisebox{-3ex}{\mbox{\includegraphics{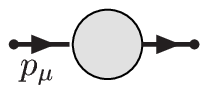}}}
\equiv
\frac{\s{p}+m}{\pi}
+ \frac{\s{p}+m}{\pi}\ \Sigma\ \frac{\s{p}+m}{\pi}
+ \ldots 
\]
\begin{equation}
= \frac{(\s{p}+m)(1-{\sigma}_2)+{\sigma}_1}
{\pi(1-{\sigma}_2)^2 - 2m{\sigma}_1(1-{\sigma}_2) -
{\sigma}^2_1}\; .
\label{eq:nucleon-2-point}
\end{equation}
To keep the pole position in the physical mass
$m$ we, therefore, need the prescription
\begin{equation}
\sigma_1 (0) = 0\ ,
\label{eq:app-RP-mass}
\end{equation}
while
$\sigma_2$ does not affect the mass shift. Further, once the condition
(\ref{eq:app-RP-mass}) holds, the full propagator near the pole is:
\[
\left.
\raisebox{-3ex}{\mbox{\includegraphics{sigma-small.eps}}}
\right|_{\scriptscriptstyle p_0 \sim \sqrt{{\bf p}^2 + m^2}}
\]
\begin{equation}
= \frac{ 1 }{1 - \sigma_2 (0) - 2m\sigma'_1(0)}
\frac{\s{p}+m}{\pi} + \ldots\ ,
\label{eq:app-residue}
\end{equation}
and the residue
$Z_f = 1$ is provided by the prescription
\begin{equation}
\sigma_2 (0) + 2m\sigma'_1(0) = 0\ .
\label{eq:app-RP-Z}
\end{equation}

First of all, let us show that the considered non-minimal insertions
do not affect the pole position of the propagator
(\ref{eq:nucleon-2-point}). Indeed,
for spin-%
$\frac{1}{2}$ case the only non-minimal tensor (matrix) structure is
$\s{p} - m$. Hence, by its very definition, a non-minimal (w.r.t.\
external nucleon line) vertex brings extra factor
$\pi$ or
$\s{p} - m$, so the contribution to
$\Sigma$ of the
{\bf n}-type graphs has the form:
\[
\Sigma^{\rm\bf n} =
\Sigma_{\rm s} \cdot \pi + \Sigma_{\rm t} \cdot (\s{p} - m).
\]
Besides, it is generally accepted that
$\Sigma$ is regular near
$p^2 = m^2$ (nucleon is massive and stable); at any finite loop
order it is justified by the very procedure of loop calculation. Then,
decomposing the (matrices!)
$\Sigma_{\rm s}$ and
$\Sigma_{\rm t}$ through
$\bf \hat{1}$ and
$\s{p} \pm m$ one easily shows that the last expression can always be
rewritten via some scalar (and regular in
$\pi = 0$) functions
$\sigma_s$ and
$\sigma_t$:
\[
\Sigma^{\rm\bf n} =
\pi \sigma_{\rm s}(\pi) \hat{\bf 1}
+ \sigma_{\rm t}(\pi) (\s{p} - m)\ ,
\]
so that the only contribution it may give to
$\sigma_1$ is proportional to
$\pi$ and, thereby, does not affect the condition
(\ref{eq:app-RP-mass}). In fact, only the minimal piece of the
on-shell contribution of
{\bf m}-type graphs (cf.
Eq.~(\ref{7.05})--(\ref{eq:7.05a})) can affect
$\sigma_1 (0)$, and it is set to zero by the RP
(\ref{eq:RP-mass}) in
Sec.~\ref{sec-RP3} (there is no imaginary part as the particle is
stable). That is, to keep the correct pole position one just imposes
the RP
(\ref{eq:RP-mass}) on the graphs with only minimal (resultant)
vertices and there is no need in the corresponding RP's for graphs
containing non-minimal vertices.

However, as seen from
Eq.~(\ref{eq:app-residue}), the non-minimal vertices do affect the
residue as they may contribute to
$\sigma'_1$ and
$\sigma_2$. Here we shall follow another logics. Separating the
contributions of
{\bf m}- and
{\bf n}-type graphs we can write the prescription
(\ref{eq:app-RP-Z}) as
\begin{equation}
\sigma^{\rm\bf m}_2 + 2m(\sigma^{\rm\bf m}_1)' =
- \left[\sigma^{\rm\bf n}_2 + 2m(\sigma^{\rm\bf n}_1)'\right]
\quad {\rm at} \; p^2 = m^2\ .
\label{eq:RP-Z-m-n}
\end{equation}
As usual, to fulfill it we must introduce counterterms. Working in
effective theory at some fixed loop order, we can separately adjust
counterterms for graphs of two different types, to be collectively
denoted as
$C^{\rm\bf m}_{_Z}$ and
$C^{\rm\bf n}_{_Z}$, respectively. Now, according to the reduction
theorem, the self-energy subgraphs with non-minimal vertices are not
present
{\em inside} the graph after the reduction, the only possibility is
pictured in
Fig.~\ref{fig:sigma}{\bf n}. Hence, the adjusted
$C^{\rm\bf n}_{_Z}$ can appear only on external legs replacing
corresponding self-energy insertions and are never put inside the
graph. In turn, any 2-leg insertion inside the graph is regulated by
$C^{\rm\bf m}_{_Z}$ only and not affected by
$C^{\rm\bf n}_{_Z}$. Simply speaking, the (non-minimal) RP's for
two-point graphs with non-minimal vertices
{\em decouple} from the other set of RP's. Only on external legs both
$C^{\rm\bf m}_{_Z}$ and
$C^{\rm\bf n}_{_Z}$ contribute. But, as it is seen from
Eq.~(\ref{eq:RP-Z-m-n}), whatever the
{\bf m}-type counterterms may be, their contribution on the left side
can always be compensated by
{\bf n}-type counterterms on the right side so that the equality,
and, thus, the condition
(\ref{eq:app-RP-Z}) holds for external leg propagator. One can treat
it as a manifestation of the well known fact that the wave function
renormalization constant is a redundant parameter of a theory.

Practically it allows one to set
$Z=1$ on the external legs automatically, having still a freedom in
relevant (off-shell) counterterms for resultant (sub-)graphs. Then the
LSZ formula says that we need to calculate only the resultant
(amputated) graphs without self-energy insertions on external legs
($S$-matrix resultant graphs). In particular, graphs with non-minimal
vertices can be discarded in the
$S$-matrix calculations --- that is exactly what we wanted to show.

Note also, that any additional mixing in external propagators due to
non-minimal vertices can be excluded by the same arguments; though, as
mentioned in
Sec.~\ref{sec-RP3}, the problem of mixing which unavoidably arise
whenever two-point function is discussed is beyond the scope of this
article.

Unfortunately, the above reasoning does not give a hand to avoid
non-minimal RP's fixing the off-shell counterterms for graphs built of
minimal vertices. In particular, corresponding 2-leg subgraphs will
appear and may introduce off-shell subdivergencies (the on-shell
behavior is, of course, regulated by RP~(\ref{eq:RP-mass})). Moreover,
while in conventional renormalizable theories we are basically limited
by first derivative of
$\Sigma$, in effective theory other derivatives can also diverge. In
forthcoming publication we will show that, accepting certain off-shell
asymptotic conditions for the full sums of given type graphs (e.g.
self-energy), one can avoid calculation of an infinite number of
non-minimal counterterms and work directly with finite parts of those
sums. In fact, the corresponding technique is already described in
Secs.~\ref{sec-Cauchy}--\ref{sec-RP3}.



\end{document}